\RequirePackage{fix-cm}

\documentclass{svjour3}                     

\usepackage{amsmath,amssymb}
\usepackage{graphicx}
\usepackage[numbers,sort&compress]{natbib}
\usepackage{chronosys}

\smartqed 
\journalname{Nonlinear Dynamics}

\def\pct{\%}
\DeclareMathOperator{\const}{const}
\DeclareMathOperator*{\sign}{sign}%
\DeclareMathOperator{\Id}{Id}%
\DeclareMathOperator{\e}{e}%

\renewcommand{\d}[1]{\ensuremath{\operatorname{d}\!{#1}}}
\newcommand{\D}[1]{\ensuremath{\operatorname{D}\!{#1}}}

\DeclareMathOperator{\spec}{spec}
\DeclareMathOperator*{\argmax}{arg\,max}

\DeclareMathOperator{\Rep}{Re_p}
\DeclareMathOperator{\neigh}{neighbor}

\begin{document}

\title{Nonlinear dynamics of inertial particles in the ocean: From
drifters and floats to marine debris and \emph{Sargassum}}
\titlerunning{Inertial ocean nonlinear dynamics} \author{Francisco
J.\ Beron-Vera} \authorrunning{Beron-Vera} \institute{F.J.\ Beron-Vera
\at Department of Atmospheric Sciences \\ Rosenstiel School of
Marine \& Atmospheric Science\\ University of Miami\\ Miami, FL
3349 USA\\ \email{fberon@rsmas.miami.edu} } \date{Received: \today/Accepted:
October 22, 2020. DOI:10.1007/s11071-020-06053-z.} 
\maketitle
	
\begin{abstract}
  Buoyant, finite-size or \emph{inertial} particle motion is
  fundamentally unlike neutrally buoyant, infinitesimally small or
  Lagrangian particle motion.  The de-jure fluid mechanics framework
  for the description of inertial particle dynamics is provided by
  the \emph{Maxey--Riley equation}.  Derived from first principles---a
  result of over a century of research since the pioneering work
  by Sir George Stokes---the Maxey--Riley equation is a Newton-type-law
  with several forces including (mainly) flow, added mass, shear-induced
  lift, and drag forces.  In this paper we present an overview of
  recent efforts to port the Maxey--Riley framework to oceanography.
  These involved: 1) including the Coriolis force, which was found
  to explain behavior of submerged floats near mesoscale eddies;
  2) accounting for the combined effects of ocean current and wind
  drag on inertial particles floating at the air--sea interface,
  which helped understand the formation of great garbage patches
  and the role of anticyclonic eddies as plastic debris traps; and
  3) incorporating elastic forces, which are needed to simulate the
  drift of pelagic \emph{Sargassum}.  Insight on the nonlinear
  dynamics of inertial particles in every case was possible to be
  achieved by investigating long-time asymptotic behavior in the
  various Maxey--Riley equation forms, which represent singular
  perturbation problems involving slow and fast variables.

\end{abstract}

\tableofcontents

\section{Introduction}

The fluid mechanics community has long observed that finite-size,
buoyant or \emph{inertial} particle motion is unlike infinitesimally
small, neutrally buoyant or Lagrangian particle motion
\cite{Michaelides-97, Cartwright-etal-10}.  However, it was not
until the seminal work of Maxey and Riley \cite{Maxey-Riley-83}
that first principles foundation was established for this observation,
representing the result of many years of research starting with the
pioneering work by Sir George Stokes in the mid 1800s \cite{Stokes-51}.
Despite the \emph{Maxey--Riley equation} provides the de-jure
framework for the study of inertial particle motion, this is only
well accepted by the fluid mechanics community \cite{Cartwright-etal-10}.
Indeed, efforts by the \emph{geophysical} fluid dynamics community
to adopt the Maxey--Riley framework are scant, including literally
a handful of applications in meteorology \cite{Provenzale-etal-98,
Provenzale-99, Dvorkin-etal-01, Sapsis-Haller-09, Haszpra-Tel-11}
and oceanography \cite{Tanga-Provenzale-94, Rubin-etal-95,
Squires-Yamazaki-95, Beron-etal-15, Haller-etal-16, Woodward-etal-19,
Aksamit-etal-20}.

The transferability of the Maxey--Riley equation to oceanography
has been hindered by the challenging problem of accounting for the
combined effects of ocean currents \emph{and} winds on particle
drift.  This problem, which at present is approached in a largely
piecemeal ad-hoc manner \cite{vanSebille-etal-18},  was addressed
recently by Beron-Vera, Olascoaga and Miron \citep{Beron-etal-19-PoF},
who derived from the Maxey--Riley equation a new equation---referred
to herein as the \emph{BOM equation}---for the drift of inertial
particles \emph{floating at the air--sea interface}.

As with the Maxey--Riley equation, the positions of the particles
in the BOM equation evolve slowly in time while their velocities
vary rapidly. This makes the BOM equation a singular perturbation
problem.  Geometric singular perturbation theory \cite{Fenichel-79,
Jones-95, Haller-Sapsis-08} can then be applied to study the long-time
asymptotic nonlinear dynamics of inertial particles on the ``slow
manifold,'' which attracts all the solutions of the BOM equation
exponentially fast in time.

\begin{figure}[t!]
  \centering%
  \includegraphics[width=\textwidth]{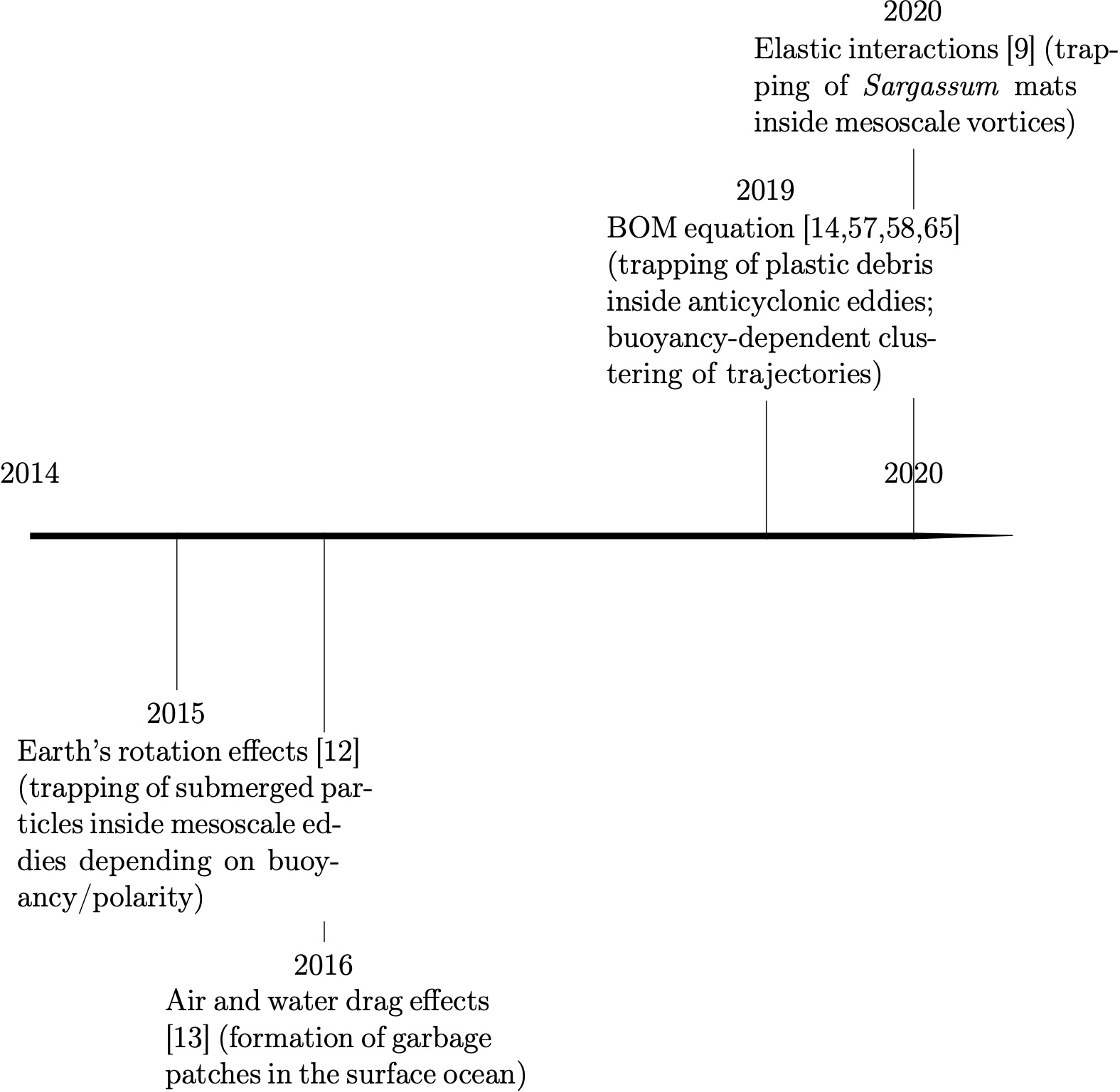}%
  \caption{Timeline of oceanographic extensions of the Maxey--Riley equation.}
  \label{fig:time}%
\end{figure}

This paper is dedicated to provide an overview of efforts leading
to the derivation of the BOM equation and of several applications
of the latter and related models in oceanographic problems (Fig.\
\ref{fig:time}).  These include the interpretation of ``Lagrangian''
observations acquired by surface and submerged drifting buoys, and
the understanding of the motion of floating marine debris and
macroalgae such as \emph{Sargassum}. Insight into the nonlinear
dynamics in every case was gained by investigating them on the
corresponding slow manifold.

The overview starts with a review of the original Maxey--Riley
equation (Sec.\ 2).  This is followed by a review of a geophysical
adaptation and a concrete oceanographic application (Sec.\ 3).  The
BOM equation is reviewed in Sec.\ 4, which includes results from
field and laboratory experiments in support of its validity.  Section
5 is dedicated to review an extension of the BOM equation to model
the motion of elastic networks of floating inertial particles that
emulate rafts of \emph{Sargassum}.  Concluding remarks on aspects
that still need to be addressed to expand the applicability of the
Maxey--Riley framework to oceanography are finally made in Sec.\
6.

\section{The original Maxey--Riley equation}

As already noted, the study of the motion of \emph{inertial} (i.e.,
buoyant, finite-size) particles was pioneered by Sir George Stokes
\cite{Stokes-51}, who solved the linearized Navier--Stokes equations
for the oscillatory motion of a small solid sphere (pendulum)
immersed in a fluid at rest. This was followed by the efforts of
\citet{Basset-88}, \citet{Boussinesq-85}, and \citet{Oseen-27} to
model a solid sphere settling under gravity, also in a quiescent
fluid.  \citet{Tchen-47} extended these efforts to model motion in
nonuniform unsteady flow by writing the resulting equation, known
as the BBO equation, on a frame of reference moving with the fluid.
Several corrections to the precise form of the forces exerted on
the particle due to the solid--fluid interaction were made along
the years \citep[e.g.,][]{Corrsin-Lumely-56}.  The now widely
accepted form of the forces was derived by \citet{Maxey-Riley-83}
from first principles, following an approach introduced by
\citet{Riley-71}.  The resulting equation, with a correction made
by \citet{Auton-etal-88}, is widely referred to as the \emph{Maxey--Riley
equation}. A similar equation was derived, independently and nearly
simultaneously, by \citet{Gatignol-83}.  \citet{Michaelides-97} and
\citet{Cartwright-etal-10} review the Maxey--Riley equation in some
detail.

\subsection{Setup}

Let $x = (x^1,x^2)$ be position on an open domain of $\mathbb R^2$.
As our ultimate interest is in geophysical applications, this domain
is actually assumed to lie on a \emph{horizontal} plane, i.e.,
perpendicular to the local gravity direction. Let $t\in\mathbb R$
be time. Let $v_\mathrm{f}(x,t)$ be the velocity of a fluid of
constant density $\rho_\mathrm{f}$ and dynamic viscosity $\mu_\mathrm{f}$.
Consider a solid sphere immersed in the fluid.  Let $a$ be its
radius, which is assumed to be small compared to any relevant length
scales of the problem, and $\rho_\mathrm{p} = \const$ its density.
Let
\begin{equation}
  \delta : = \frac{\rho_\mathrm{f}}{\rho_\mathrm{p}},
\end{equation} 
which will be referred to as \emph{buoyancy}. Indeed, particles
that are \emph{lighter} (resp., \emph{heavier}) than the carrying
fluid are characterized by $\delta > 1$ (resp., $\delta < 1$). We
will restrict attention in this section to the case $\delta \approx
1$, so the vertical motion of the particle can be neglected (rapid,
small-scale three-dimensional motions that can alter this balance
are not considered here as the interest relies on slow, large-scale
geophysical flow motions, which are essentially two dimensional).

\subsection{Forces and the resulting equation}

The Maxey--Riley equation is a classical mechanics Newton's 2nd law
with several forces describing the motion of a small solid sphere
immersed in the unsteady nonuniform flow of a homogeneous, viscous
fluid.  As such, it represents an ordinary differential equation
that provides an approximation to the motion of inertial particles.
Its exact motion is controlled by the Navier--Stokes equation with
moving boundaries.  This is necessarily described by partial
differential equations, which much more difficult to solve and
analyze than an ordinary differential equation.

The Maxey--Riley equation includes several forcing terms which
prevent inertial particles from adapting their velocities to
instantaneous changes in the carrying flow field.  Normalized by
particle mass $m_\mathrm{p} = \frac{4}{3}\pi a^3\rho_\mathrm{p}$
the relevant forces for the \emph{horizontal} motion are:
\begin{enumerate}
  \item the \emph{flow force} exerted on the particle by the
  undisturbed fluid:
  \begin{equation}
    F_\mathrm{flow} =
    \frac{m_\mathrm{f}}{m_\mathrm{p}}\frac{\D{v_\mathrm{f}}}{\D{t}}
	 \label{eq:F_flow}
  \end{equation}
  where $m_\mathrm{f} = \frac{4}{3}\pi a^3\rho_\mathrm{f}$ is the
  mass of the displaced fluid and $\smash{\frac{\D{}}{\D{t}}}v_\mathrm{f}$
  is the fluid velocity's material derivative, namely,
  $\smash{\frac{\D{}}{\D{t}}} v_\mathrm{f} =
  \smash{\big[\frac{\d{}}{\d{t}}v_\mathrm{f}(x,t)\big]_{x =
  X_\mathrm{f}(t)}} = \partial_t v_\mathrm{f} + (\nabla
  v_\mathrm{f})v_\mathrm{f}$, where $x = X_\mathrm{f}(t)$ is a fluid
  trajectory; \item the \emph{added mass force} resulting from part
  of the fluid moving with the particle:
  \begin{equation}
    F_\mathrm{mass} =
    \frac{\frac{1}{2}m_\mathrm{f}}{m_\mathrm{p}}\left(\frac{\D{v_\mathrm{f}}}{\D{t}}
    - \dot v_\mathrm{p}\right)
  \end{equation}
  where $\dot v_\mathrm{p}$ is the acceleration of an inertial
  particle with trajectory $x = X_\mathrm{p}(t)$, i.e., $\dot
  v_\mathrm{p} =
  \smash{\frac{\d{}}{\d{t}}\left[v_\mathrm{p}(x,t)\right]_{x=X_\mathrm{p}(t)}}
  = \partial_t v_\mathrm{p}$ where $v_\mathrm{p} = \partial_t
  X_\mathrm{p} = \dot x$ is the inertial particle velocity; \item
  the \emph{lift force}, which arises when the particle rotates as
  it moves in a (horizontally) sheared flow,
  \begin{equation}
	 F_\mathrm{lift} =
	 \frac{\frac{1}{2}m_\mathrm{f}}{m_\mathrm{p}}\omega_\mathrm{f}
	 J(v_\mathrm{f} - v_\mathrm{p}),
  \end{equation}
  where $\omega_\mathrm{f} = \partial_1 v^2_\mathrm{f} - \partial_2
  v^1_\mathrm{f}$ is the (vertical) vorticity of the fluid and
  \begin{equation}
	 J := 
	 \begin{pmatrix}
		0 & -1\\
		1 &  0
	 \end{pmatrix};
  \end{equation}
  \item the \emph{drag force} caused by the fluid viscosity,
  \begin{equation}
	 F_\mathrm{drag} = \frac{12\mu_\mathrm{f}
	 \frac{A_\mathrm{f}}{\ell_\mathrm{f}}}{m_\mathrm{p}}
	 (v_\mathrm{f} - v_\mathrm{p})
	 \label{eq:F_fdrag}
  \end{equation}
  where $A_\mathrm{f}$ ($=\pi a^2$) is the projected area of the
  particle and $\ell_\mathrm{f}$ ($=2a$) is the characteristic
  projected length. 
\end{enumerate}
The Maxey--Riley equation, $\dot v_\mathrm{p} = F_\mathrm{flow} +
F_\mathrm{mass} + F_\mathrm{lift} + F_\mathrm{drag}$, reads
\begin{equation}
  \dot v_\mathrm{p} +
  \left(\tfrac{1}{2}R\omega_\mathrm{f} J + \frac{\Id}{\tau}\right)
  v_\mathrm{p} = \tfrac{3}{2}R\frac{\D{v_\mathrm{f}}}{\D{t}} +
  \left(\tfrac{1}{2}R\omega_\mathrm{f} J + \frac{\Id}{\tau}\right) v_\mathrm{f},
  \label{eq:MR83}
\end{equation}
where
\begin{equation}
  R := \frac{2\delta}{2+\delta},\quad \tau :=
  \tfrac{2}{3}R^{-1}\cdot
  \frac{a^2\rho_\mathrm{f}}{3\mu_\mathrm{f}}.
\end{equation}
Here $\tau$ is the \emph{inertial particle's response time to the
medium} or \emph{Stokes' time}. Note that $0 \leq R < 2$ with $R >
\smash{\frac{2}{3}}$ (resp., $R < \smash{\frac{2}{3}}$) characterize
light (resp., heavy) particles.

\begin{remark}
A pertinent question is whether the fluid particle equation $\dot
x = v_\mathrm{f}$ is recovered from \eqref{eq:MR83} when $R =
\smash{\frac{2}{3}}$ and $\tau = 0$.   The answer to this question
is affirmative, essentially.  Yet it requires elaboration, provided
in the next section, as \eqref{eq:MR83} becomes \emph{singular} at
$\tau = 0$.
\end{remark}

\begin{remark}
Except for the lift force, due to \citet{Auton-87}, the forces just
described are included in the paper by \citet{Maxey-Riley-83}, yet
with a different form of the added mass term, which corresponds to
the correction due to \citet{Auton-etal-88}.  The particular form
of the lift force above is found in \citep[][Ch.\ 4]{Montabone-02};
similar forms are considered in \citet{Henderson-etal-07} and
\citet{Sapsis-etal-11}. A condition for the validity of the lift
force is \citep{Auton-87, Auton-etal-88}
\begin{equation}
  \frac{|\omega_\mathrm{f}|a}{|v_\mathrm{f} - v_\mathrm{p}|} \ll 1.
\end{equation}
\end{remark}

\begin{remark}
The Maxey--Riley equation \eqref{eq:MR83} was derived under the
assumption that the \emph{particle Reynolds number}
\begin{equation}
  \Rep := \frac{V_\mathrm{slip}\cdot
  \ell_\mathrm{f}}{\mu_\mathrm{f}/\rho_\mathrm{f}} \ll 1,
  \label{eq:rep}
\end{equation}
where $V_\mathrm{slip}$ is a measure of the \emph{difference} between
$v_\mathrm{p}$ and $v_\mathrm{f}$. We will see that this
is indeed well satisfied for sufficiently small particles since in
that case $v_\mathrm{p}$ is asymptotically close to $v_\mathrm{f}$.
\end{remark}

\begin{remark}
The general form of the drag force (e.g., \cite{Kundu-etal-12})
per unit particle mass is:
\begin{equation}
  F_\mathrm{drag} =
  \frac{\tfrac{1}{2}\rho_\mathrm{f}C_\mathrm{D}A_\mathrm{f}}{m_\mathrm{p}}
  |v_\mathrm{f} - v_\mathrm{p}|(v_\mathrm{f} -
  v_\mathrm{p}).
  \label{eq:CD}
\end{equation}
A particle in a flow in Stokes' regime, for which $\mathrm{Re_p} <
1$, is characterized by a drag coefficient of the form
\begin{equation}
  C_\mathrm{D} = \frac{24}{\mathrm{Re_p}}.
\end{equation}
Plugging  $C_\mathrm{D} = 24/\mathrm{Re_p}$ in the general drag
formula \eqref{eq:CD}, one gets, upon setting $V_\mathrm{slip} =
|v_\mathrm{f} - v_\mathrm{p}|$,
\begin{equation}
  F_\mathrm{drag} 
  =
  \frac{\tfrac{1}{2}\rho_\mathrm{f}\frac{24}{\mathrm{Re_p}}A_\mathrm{f}}{m_\mathrm{p}}
  |v_\mathrm{f} - v_\mathrm{p}|(v_\mathrm{f} -
  v_\mathrm{p}) 
  =
  \frac{12\mu_\mathrm{f}\frac{A_\mathrm{f}}{\ell_\mathrm{f}}}{m_\mathrm{p}}(v_\mathrm{f}
  - v_\mathrm{p}).
\end{equation}
\end{remark}

\begin{remark}
In writing the Maxey--Riley equation \eqref{eq:MR83} we have ignored
the Basset--Boussinesq history or memory term, which is an integral
term that makes the equation a fractional differential equation
\cite{Daitche-Tel-11, Daitche-Tel-14, Langlois-etal-15}. This may
be (has been) neglected under low recurrence time grounds
\cite{Sudharsan-etal-16}.  It has been also noted \cite{Daitche-Tel-11}
that it mainly tends to slow down the inertial particle motion
without changing its qualitative dynamics fundamentally.  However,
the effects of the memory terms remain the subject of active research
\cite{Olivier-etal-14, Prasath-etal-18, Haller-19}.  We have also
ignored so-called Faxen corrections (terms of the form
$a^2\nabla^2v_\mathrm{f}$) in the added mass and drag forces; this
is much easier to justify.
\end{remark}

\subsection{Slow manifold reduction}

Because of the small-particle-size assumption involved in the
derivation of the Maxey--Riley equation, it is natural to investigate
its asymptotic behavior when $\tau = O(\varepsilon)$ as $\varepsilon\to
0$, where $0 \le \varepsilon \ll 1$ is a parameter that we will use
to measure smallness (of any nature) throughout this paper (in this
case it can be interpreted as a Stokes number \citep{Haller-Sapsis-08}).
In this limit, the Maxey--Riley equation \eqref{eq:MR83} involves
a \emph{fast} variable, $v_\mathrm{p}$, changing at $O(\varepsilon^{-1})$
speed, and a \emph{slow} variable, $x$, changing at $O(1)$ speed,
which makes \eqref{eq:MR83} a \emph{singular} perturbation problem.
This can be seen by putting \eqref{eq:MR83} in \emph{system} form,
viz.,
\begin{equation}
	 \dot x = v_\mathrm{p},\quad \dot v_\mathrm{p} =
	 \frac{v_\mathrm{f}-v_\mathrm{p}}{\tau} +
	 \tfrac{3}{2}R\frac{\D{v_\mathrm{f}}}{\D{t}} +
	 \tfrac{1}{2}R\omega_\mathrm{f}J\left(v_\mathrm{f}-v_\mathrm{p}\right),\quad
    \dot t = 1.
  \label{eq:slow}
\end{equation}
Changing $t$ by the \emph{fast} time $s = \frac{t -
t_0}{\varepsilon}$ \cite{Haller-Sapsis-08}, system \eqref{eq:slow}
recasts as
\begin{equation}
	 x' = \varepsilon v_\mathrm{p},\quad v'_\mathrm{p} =
	 \frac{v_\mathrm{f}-v_\mathrm{p}}{\tau/\varepsilon} +
	 \tfrac{3}{2}R\frac{\D{v_\mathrm{f}}}{\D{t}} +
	 \tfrac{1}{2}R\omega_\mathrm{f}J\left(v_\mathrm{f}-v_\mathrm{p}\right),\quad
    t' = \varepsilon,
  \label{eq:fast}
\end{equation}
where $' = \frac{\d{}}{\d{s}}$.  There are two distinguished limiting
behaviors for the above systems.  Setting $\tau = 0$ in the
\emph{fast} system \eqref{eq:fast},
\begin{equation}
	 x' = 0,\quad v'_\mathrm{p} =
	 v_\mathrm{f}-v_\mathrm{p},\quad
    t' = 0,
  \label{eq:fast0}
\end{equation}
from which one obtains that $x$ and $t$ do not change, yet the
motion is accelerated. This physically absurd situation however is
consistent with $v_\mathrm{p}$ being the fast variable and $x$ and
$t$ the slow variables.  The corresponding limit of the \emph{slow}
system \eqref{eq:slow},
\begin{equation}
	 \dot x = v_\mathrm{p},\quad 0 =
	 v_\mathrm{f}-v_\mathrm{p},\quad
    \dot t = 1,
  \label{eq:slow0}
\end{equation}
gives the motion on
\begin{equation}
  M_0 : = \big\{(x,v_\mathrm{p},t) : v_\mathrm{p} = v_\mathrm{f}(x,t)\big\},
  \label{eq:M0}
\end{equation}
which is the set of \emph{equilibria} of \eqref{eq:fast0}.  Thus
while \eqref{eq:fast0} has a large set of equilibria on which the
motion is trivial, \eqref{eq:slow0} blows the flow on this set up
to produce nontrivial behavior, yet leaving the flow \emph{off} the
set \emph{undetermined}. This makes the resolution of \eqref{eq:slow}
a singular perturbation problem.

The goal of the \emph{geometric singular perturbation theory}
(\emph{GSPT}) of \citet{Fenichel-79} (cf.\ the lecture notes of
\citet{Jones-95} for additional insight), extended to
nonautonomous systems by \citet{Haller-Sapsis-08}, is to capture
the fast and slow aspects of the motion in systems like \eqref{eq:slow}
simultaneously. This is accomplished in the case of \eqref{eq:slow}
by examining the motion for $\tau = O(\varepsilon)$ as $\varepsilon\to
0$ as follows.

Assume that $v_\mathrm{f}(x,t)$ is smooth in each of its arguments.
Then $M_0$ represents a 3-dimensional, invariant, globally attracting,
normally hyperbolic manifold\footnote{If $t\in [t_1,t_2]\subset\mathbb
R$, as in applications involving measurements, $M_0$ will not form,
strictly speaking, a manifold since it will necessary include
corners.  Yet $M_0\!\setminus\!\partial M_0$ represents a well-defined
manifold.} for \eqref{eq:fast0} \cite{Fenichel-71}.  Indeed, $M_0$
is filled with equilibria of \eqref{eq:fast0}, whose linearization
at each point on $M_0$ has 3 null eigenvalues, with corresponding
neutral eigenvectors tangent to $M_0$, and 2 eigenvalues equal
to $-1$, with corresponding contracting eigenvectors such that
$(x,t) = \const$. Since tangencies are ruled out by the smoothness
assumption on $v_\mathrm{f}$, this guarantees that contraction
occurs in the normal direction to $M_0$ exclusively.  More explicitly,
integrating \eqref{eq:fast0},
\begin{equation}
  x = x_0,\quad v_\mathrm{p} = v_\mathrm{f}(x_0,0) +
  \big(v_\mathrm{p}(0) - v_\mathrm{f}(x_0,0)\big)
  \cdot\e^{-s},\quad t = t_0,
\end{equation}
which shows that any initial condition of \eqref{eq:fast0} has its
$\omega$-limit in $M_0$ and that the normal projection of a normal
perturbation to $M_0$ decays under the (linearized) flow as $\e^{-s}$,
while the tangential projection grows as $1 - \e^{-s}$, except at
critical points where it vanishes.

\begin{figure}[t!]
  \centering%
  \includegraphics[width=\textwidth]{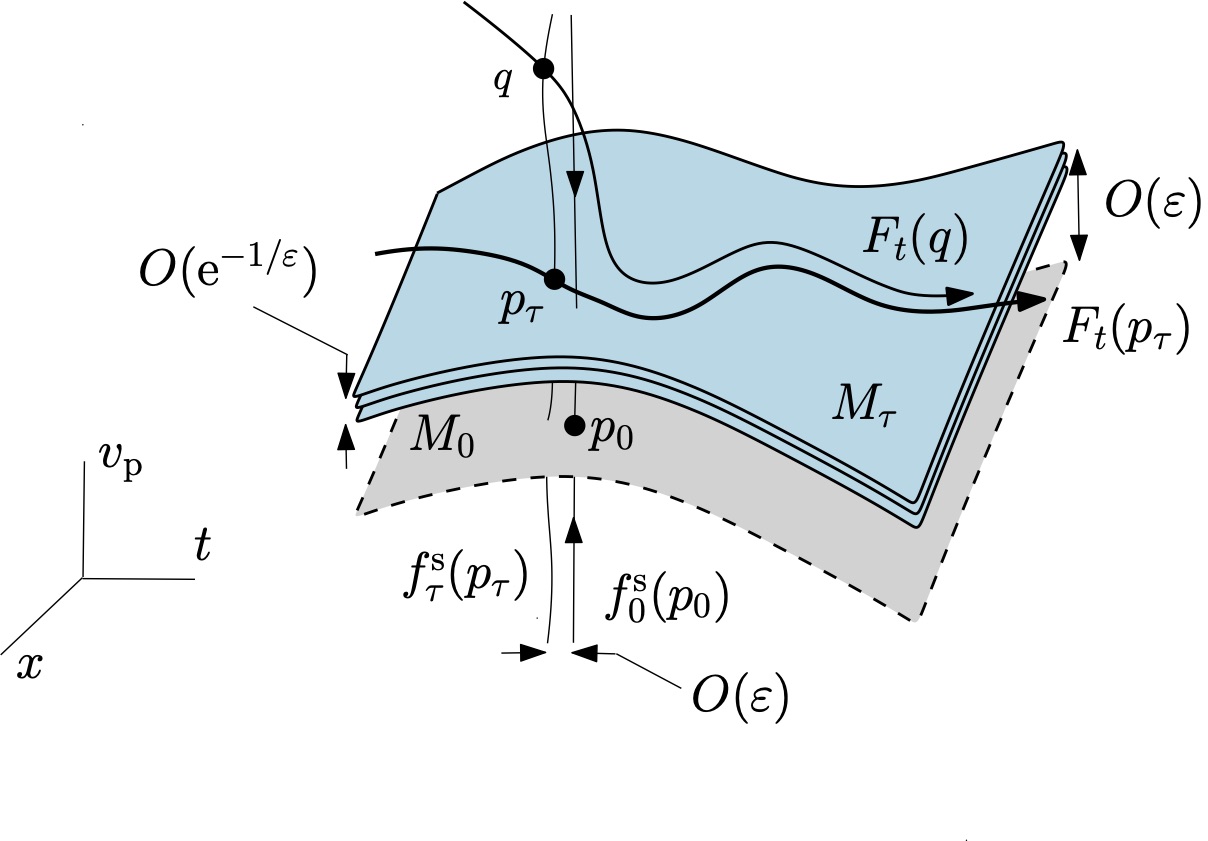}%
  \caption{Geometry of the Maxey--Riley equation \eqref{eq:MR83}.
  In the $\tau = 0$ limit the equation has a globally attracting,
  normally hyperbolic invariant manifold, $M_0$, filled with
  equilibria.  For $\tau = O(\varepsilon)$ as $\varepsilon\to 0$,
  there exists a unique up to an error of
  $O(\smash{\mathrm{e}^{-1/\varepsilon}})$, locally invariant,
  globally attracting manifold $M_\tau$, which lies at an
  $O(\varepsilon)$ distance to $M_0$ and is diffeomorphic to it.
  Restricted to $M_\tau$, the Maxey--Riley equation slowly varies
  while controlling the motion off $M_\tau$ as follows. When $\tau
  = 0$, each point off $M_0$ belongs to the stable manifold of
  $M_0$, which is foliated by its distinct stable fibers or stable
  manifolds of points $p_0$ on $M_0$, $f_0^s(p_0)$, satisfying
  $(x,t) = \const$.  The stable manifold of $M_0$ and its stable
  fibers perturb along with $M_0$.  Consequently, for $\tau =
  O(\varepsilon)$ each point $q$ off the slow manifold $M_\tau$ is
  connected to a point $p_\tau$ on $M_\tau$ by a fiber $f_\tau^s(p_\tau)$
  in the sense that it follows a trajectory $F_t(q)$ that converges
  to its partner $F_t(p_\tau)$ on $M_\tau$ exponentially fast in
  time.}
  \label{fig:gspt}%
\end{figure}

Then nonautonomous GSPT guarantees the existence of a locally
invariant (i.e., up to trajectories leaving through the boundary),
globally attracting, normally hyperbolic manifold
\begin{equation}
  M_\tau := \Big\{(x,v_\mathrm{p},t) : v_\mathrm{p} = v_\mathrm{f}(x,t)
  + \sum_1^r \tau^n v_n(x,t) + O(\varepsilon^{r+1})\Big\} 
  \label{eq:Mtau}
\end{equation}
for \eqref{eq:slow}---or \eqref{eq:fast}---when $\tau = O(\varepsilon)$
as $\varepsilon\to 0$, called a \emph{slow manifold}, which is
$O(\varepsilon)$-close to the \emph{critical manifold} $M_0$ and
$C^r$-diffeomorphic to it for any $r < \infty$.  Restricted to
$M_\tau$, \eqref{eq:slow} slowly varies while controlling the motion
off $M_\tau$ as follows.  When $\tau \propto \varepsilon = 0$, each
point off $M_0$ belongs to the stable manifold of $M_0$, which is
foliated by its distinct stable fibers (stable manifolds of points
on $M_0$) satisfying $(x,t) = \const$.  The stable manifold of $M_0$
and its stable fibers perturb along with $M_0$.  As a result, for
$\tau = O(\varepsilon)$ as $\varepsilon\to 0$ each point off the
slow manifold $M_\tau$ is connected to a point on $M_\tau$ by a
fiber in the sense that it follows a trajectory that approaches its
partner on $M_\tau$ exponentially fast in time.  The geometry of
these results is illustrated in Fig.\ \ref{fig:gspt}.

The functions $v_n(x,t)$ that determine $M_\tau$ are obtained by
substituting the asymptotic expansion in \eqref{eq:Mtau} in the
second equation of \eqref{eq:slow} with the first equation in mind,
which tells that $\dot \varphi = \partial_t \varphi + (\nabla
\varphi)v_\mathrm{p}$ for any function $\varphi(x,t)$, and then
equating terms of like order of $\varepsilon$.  This gives \citep[cf.\
App.\ C of Ref.][]{Beron-etal-19-PoF}
\begin{align}
  v_1 &= \big(\tfrac{3}{2}R-1\big)\frac{\D{v_\mathrm{f}}}{\D{t}}\\ v_n &= -
  \tfrac{1}{2}R\omega Jv_{n-1} - \frac{\D{}}{\D{t}} v_{n-1} - (\nabla
  v)v_{n-1} - \sum_{m=1}^{n-2}(\nabla v_m)v_{n-m-1},\quad
  n \ge 2.\label{eq:vn}
\end{align}
One then finds that 
\begin{equation}
  \dot x = v_\mathrm{p} = v_\mathrm{f} +
  \tau\big(\tfrac{3}{2}R-1\big)\frac{\D{v_\mathrm{f}}}{\D{t}}
  \label{eq:MRslow}
\end{equation}
describes, with an $O(\varepsilon^2)$ error, the asymptotic dynamics
of the Maxey--Riley equation on the slow manifold $M_\tau$ in the
form of a \emph{regular perturbation problem}.

We will refer \eqref{eq:MRslow} to as the \emph{reduced Maxey--Riley
equation}.  This reduced equation coincides with that derived in
\cite{Haller-Sapsis-08} ignoring the lift term, which makes a
higher-order contribution in $\varepsilon$ to $M_\tau$---it appears
at $O(\varepsilon^2)$; cf.\ \eqref{eq:vn}.  The special case of a
steady cellular carrying flow, also without lift term, was considered
by \citet{Rubin-etal-95}, who first reported an analysis of the
Maxey--Riley equation using (autonomous) GSPT.

\begin{remark}
The slow manifold is not unique.  There typically is a family of
slow manifolds with members lying at an
$\smash{O(\mathrm{e}^{-1/\varepsilon})}$ distance from one another
\cite{Fenichel-79}.  On the other hand, rapid changes in the carrying
fluid velocity $v_\mathrm{f}$ will lead to rapid changes in $M_\tau$,
thereby restricting its ability to absorb solutions over finite
time.  In other words, the convergence to $M_\tau$ may not be
monotone \cite{Haller-Sapsis-08}.  \label{re:mono}
\end{remark}

\begin{remark}
As a 2-dimensional system, the reduced Maxey--Riley equation
\eqref{eq:MRslow} is numerically less expensive to solve than the
full Maxey--Riley equation \eqref{eq:MR83}, which is 4-dimensional.
As such, it requires specification of initial positions only, rather
than initial positions and velocities, which are generally not
available.  Also, unlike the full equation, the reduced equation
is not subjected to numerical instability in backward time integration
\cite{Haller-Sapsis-08}, which is useful in source inversion.
Furthermore, as we will see, it provides insight that is difficult
to be attained using the full Maxey--Riley equation.
\end{remark}

\begin{remark}
The slow manifold $M_\tau$ \eqref{eq:Mtau} and the restriction of
the Maxey--Riley equation to $M_\tau$ \eqref{eq:MRslow} formally
satisfy the definition of \emph{inertial manifold} and \emph{inertial
equation}, respectively, developed for the study of attractors in
infinite-dimensional dynamical systems \cite{Temam-90}.  In such
systems, actual attractors are hard to compute and are generally
not even manifolds.  The inertial manifold is easier to compute,
smooth, and contains the attractor.  It is unclear to the author
why the term ``inertial'' is used to denote such manifolds.  Here
``inertia'' refers to resistance of an object to a change in its
velocity.
\end{remark}

\subsection{Neutrally buoyant particles}\label{sec:neutral}

The neutrally buoyant case $\delta = 1$ or, equivalently, $R =
\smash{\frac{2}{3}}$ deserves a separate discussion.  The results
above imply that neutrally buoyant particle motion synchronizes
exponentially fast with Lagrangian particle motion when particles
are sufficiently small, i.e., $\tau = O(\varepsilon)$.  Indeed, on
the slow manifold $\dot x = v_\mathrm{p} = v_\mathrm{f}$ with an
$O(\varepsilon^2)$ error.  However, \citet{Babiano-etal-00} noted
that, in the case with no lift force, the manifold
\begin{equation}
  N := \big\{(x,v_\mathrm{p},t) : v_\mathrm{p} = v_\mathrm{f}(x,t)\big\},
  \label{eq:N}
\end{equation}
while invariant for any $\tau$, may become unstable for $\tau$
large.  The \emph{neutral manifold} $N$ coincides with the critical
manifold $M_0$ in \eqref{eq:M0}.  But this is not true in the ocean
adaption(s) of the Maxey--Riley equation discussed below, so is
appropriate to use different labels for these two manifolds.

The invariance of $N$ holds even with lift term present \citep[][App.\
B]{Beron-etal-19-PoF}, as one finds by writing the Maxey--Riley
equation \eqref{eq:MR83} with $R = \smash{\frac{2}{3}}$ following
\cite{Babiano-etal-00} as
\begin{equation}
   \dot y = A y,\quad y := v_\mathrm{p} - v_\mathrm{f}\quad A : = - \big(\nabla
	v_\mathrm{f} + \tfrac{1}{3}\omega_\mathrm{f} J + \tau^{-1}\Id\big).
	\label{eq:doty}
\end{equation}
Here $\dot v_\mathrm{f} = \partial_t v_\mathrm{f} + (\nabla
v_\mathrm{f})v_\mathrm{p}$ is the total derivative of $v$ taken along
an inertial particle trajectory, satisfying $\dot x = v_\mathrm{p}$.
Clearly $y = 0$ trivially solves $\dot y = A y$.  From this it
follows that $N$ is invariant for any $\tau$.

The possibility of growing perturbations off $N$ follows from
inspecting the sign of the \emph{instantaneous stability indicator},
discussed by \citet{Sapsis-Haller-08},
\begin{equation}
  \Lambda (x_0,t_0) := \lim_{t\to t_0}
  \frac{2}{t-t_0}\log||P_{t_0}^t||_2,
  \label{eq:isi}
\end{equation}
where $P_{t_0}^t$ is the fundamental matrix solution of \eqref{eq:doty}.
Taylor expanding $P_{t_0}^t$,
\begin{equation}
  (||P_{t_0}^t||_2)^2 = 1 - 2\min\spec(S(x_0,t_0)+\tau^{-1}\Id)\cdot (t-t_0)
  + O((t-t_0)^2),
\end{equation}
where $S := \smash{\frac{\nabla v + (\nabla v)^\top}{2}}$ is the
strain-rate tensor, and $\log\left(1+\smash{\sum_1^\infty}
c_n\varepsilon^n\right) = c_1\varepsilon + O(\varepsilon^2)$, it
follows that
\begin{equation}
  \Lambda(x_0,t_0) = -2\min\spec(S(x_0,t_0) + \tau^{-1}\Id)
\end{equation}
\citep[cf.\ App.\ B of Ref.][for deatils]{Beron-etal-19-PoF}.
Replacing $(x_0,t_0)$ with $(x(t),t)$ one concludes that
\emph{instantaneous} divergence away from $N$ will \emph{necessarily}
take place \emph{where}
\begin{equation}
  \tau > \frac{2}{\sqrt{S_\mathrm{n}^2 +
  S_\mathrm{s}^2} - \nabla\cdot v}.
  \label{eq:Ndiv}
\end{equation}
Here $S_\mathrm{n} := \partial_1v_\mathrm{f}^1 - \partial_2v_\mathrm{f}^2$
and $S_\mathrm{s} := \partial_2v_\mathrm{f}^1 + \partial_1v_\mathrm{f}^2$
are normal and shear strain components, respectively.  Condition
\eqref{eq:Ndiv} reduces to $\tau > 1/\sqrt{|\det S|}$ in the
(geophysically relevant) incompressible case $\nabla\cdot v = 0$.
This coincides with that one obtained in \cite{Sapsis-Haller-08}
ignoring the lift force, which is seen to play no role in setting
the local instability of $N$.

A \emph{sufficient} condition for \emph{global} attractivity on $N$ is
provided by the violation of \eqref{eq:Ndiv} \emph{everywhere} as
it follows by noting that
\begin{equation}
  |y(t)| \le |y(t_0)| \mathrm{e}^{-\textstyle{\int_{t_0}^t}
  \min\spec(S(x(s),s) + \tau^{-1}\Id)\d{s}}, 
  \label{eq:Ncon}
\end{equation}
where $y^\top Ay \le \smash{\frac{1}{2}}\max\spec(A+A^\top)\cdot|y|^2
= - \min\spec\big(S + \tau^{-1}\Id\big)\cdot|y|^2$ was taken into
account \citep[][App.\ B]{Beron-etal-19-PoF}.

\begin{remark}
It is important to realize that for $(y_0,x_0)$ given, convergence
on $N$ does not imply convergence on a fluid trajectory starting
on $x_0$. Consider $v_\mathrm{f} = V = \const$.  This implies that $\omega =
0$. It readily follows that $y = y_0\smash{\mathrm{e}^{-t/\tau}}$
and $x = x_0 + V t + \tau y_0 (1 - \smash{\mathrm{e}^{-t/\tau}})$.
Note that while $y\to 0$ as $t\to \infty$, $x\to x_0 + V t + \tau
y_0$, when the fluid trajectory starting from $x_0$ is $x_0 + V t$.
Clearly, coincidence is expected only when the neutrally buoyant
particle is sufficiently small, i.e., $\tau = O(\varepsilon)$,
consistent with $N$ lying at $O(\varepsilon^2)$ distance to
$M_\tau$, which attracts all solutions in that
limit.\label{re:neutral}
\end{remark}

\begin{remark}
It turns out that \eqref{eq:Ndiv} is also a necessary condition for
the instability of perturbations off the slow manifold $M_\tau$
given in \eqref{eq:Mtau}, i.e.,  without the neutrally buoyant
particle constraint.  This follows from the local instability
analysis of an arbitrary invariant manifold developed by
\citet{Haller-Sapsis-10}.  The main result of that work is the
derivation of a local stability indicator, called \emph{normal
infinitesimal Lyapunov exponent} or \emph{NILE}, which is related
to the instantaneous stability indicator \eqref{eq:isi}.  For a
system of the form
\begin{equation}
  \dot x = f(x,y,t),\quad \dot y = g(x,y,t),
\end{equation}
the NILE for a perturbation off a general invariant manifold of graph form, 
\begin{equation}
  M = \{(x,y,t) : y = h(x,t)\},
\end{equation}
is given by 
\begin{equation}
  \sigma(x,t) = \tfrac{1}{2}\max\spec\big(\Gamma(x,t)
  + \Gamma(x,t)^\top\big) 
  \label{eq:nile}
\end{equation}
where
\begin{equation}
  \Gamma(x,t) : = (\partial_yg - \partial_xh\partial_yf)\vert_{y = h(x,t)}.
  \label{eq:sigma}
\end{equation}
In other words, the manifold $M$ becomes locally repelling in $(x,t)$
regions where the NILE is positive. For a perturbation off slow
manifold $M_\tau$ of the Maxey--Riley equation \eqref{eq:MR83},
$\sigma(x,t) = - \min\spec\smash{\big(S(x,t) + \tau^{-1}\Id\big)}
+ O(\varepsilon)$, whose lowest-order contribution is positive for
$(x,t)$ where \eqref{eq:Ndiv} holds.  It should be realized that
this result does not contradict the nonautonomous GSPT result on
the global attractivity of $M_\tau$, which is an asymptotic result.
It is consistent with Remark \ref{re:mono} on the possible nonmonotonic
convergence to $M_\tau$ \cite{Haller-Sapsis-08}.  \label{re:nile}
\end{remark}

\section{Geophysical extension of the original Maxey--Riley equation}

As a first step toward preparing for oceanographic applications of
the Maxey--Riley equation, we need to move away from the laboratory
frame, taking into account the effects of the rotation of the Earth
and its curvature.  The adaption that follows actually applies more
widely to geophysical flows, such as Earth atmospheric flows and
possibly also planetary atmospheric flows.  Indeed, this adaptation
has been suitable to provide insight into aspects of inertial motion
in the ocean \cite{Beron-etal-15,Haller-etal-16} and also in the
stratosphere \cite{Provenzale-99}.

\subsection{The Coriolis force}

Let $a_\odot$ be the Earth's radius, and consider the rescaled
longitude ($\lambda$) and latitude ($\vartheta$) coordinates:
\begin{equation}
  x^1 = (\lambda-\lambda_0)\cdot
  a_\odot\cos\vartheta_0,\quad
  x^2 = (\vartheta-\vartheta_0)\cdot a_\odot,
\end{equation}
where $(\lambda_0,\vartheta_0)$ is a reference location on the planet's
surface. Consider the following geometric coefficients
\cite{Ripa-JPO-97b}:
\begin{equation}
  \gamma_\odot := \sec\vartheta_\odot\cos\vartheta,\quad
  \tau_\odot := a^{-1}_\odot\tan\vartheta.
\end{equation}
The (horizontal) velocity of a fluid particle and its acceleration
as measured by a \emph{terrestrial} observer are
\cite{Ripa-JPO-97b,Beron-03}
\begin{equation}
v_\mathrm{f} = m_\odot\dot x,\quad m_\odot:=
\begin{pmatrix}
  \gamma_\odot & 0\\
	0 & 1
 \end{pmatrix}
 ,\quad
 a_\mathrm{f} = \dot v_\mathrm{f} + (f + \tau_\odot
 v_\mathrm{f}^1)Jv_\mathrm{f},
 \label{eq:a-f}
\end{equation}
respectively, where $f := 2\Omega\sin\vartheta$ is the Coriolis
``parameter.'' 

A very enlightening way to derive the formula for the acceleration
is from Hamilton's principle, where the Lagrangian is written by
an observer standing on a fixed frame but with the coordinates
related to those rotating with the planet.  This way the only force
acting on the particle (in the absence of any other forces) is the
gravitational one.  This is in essence what Pierre Simon de Laplace
(1749--1827) did to derive his theory of tides and at the same time
discover the Coriolis force \emph{over a quarter of a century before}
Gaspard Gustave de Coriolis (1792--1843) was born
\cite{Ripa-RMF-95,Ripa-FCE-97,Ripa-JPO-97b}.

\begin{remark}
Indeed, for an observer standing on a fixed frame, the only force
acting on a free particle on the assumed smooth, frictionless
surface, $S$, of the Earth is the gravitational force. Thus, on
$S$, we must have $V + V_\mathrm{C} = 0$ (without loss of generality)
where $V$ and $V_\mathrm{C}$ are gravitational and centrifugal
potentials, respectively \cite{Ripa-JPO-97b,Beron-03}.  The centrifugal
potential is easy to express: $V_\mathrm{C} =
\smash{\tfrac{1}{2}a_\odot^2\Omega^2\cos^2\vartheta}$.  In turn,
the kinetic energy of the particle as measured by the fixed observer,
$T := \smash{\tfrac{1}{2}a_\odot^2} (\cos^2\vartheta(\dot\vartheta
+ \Omega)^2 + \dot\vartheta^2)$.  Using Pedro Ripa's convenient
trick \cite{Ripa-JPO-97b} to augment the number of generalized
coordinates from $(x^1,y^1)$ to $(x^1,x^2,v^1 = \gamma_\odot\dot
x^1, v^2 = \dot x^2)$, the Lagrangian, $L := T- V \equiv \gamma_\odot\dot
x^1 \smash{\big(v^1 + \tfrac{1}{2}\tau_\odot^{-1}f\big)} + \dot
x^2v^2 - \smash{\tfrac{1}{2}\big((v^1)^2 + (v^2)^2\big)}$, which
leads (directly) to a motion equation equation in system form, viz.,
$\dot x  = m_\odot^{-1}v$ and $\dot v + (f + \tau_\odot v^1)Jv =
0$; cf.\ \eqref{eq:a-f}.
\end{remark}

By a similar token, the fluid's Eulerian acceleration takes the form
\begin{equation}
  \frac{\D{v_\mathrm{f}}}{\D{t}} + (f + \tau_\odot
  v_\mathrm{f}^1)Jv_\mathrm{f},
  \label{eq:DvDt-f}
\end{equation}
where
\begin{equation}
  \frac{\D{v_\mathrm{f}}}{\D{t}} = \partial_t v_\mathrm{f} + (\nabla
  v_\mathrm{f}) \dot x = \partial_t v_\mathrm{f} +
  (\gamma_\odot^{-1}\partial_1 v_\mathrm{f})v_\mathrm{f}^1 +
  (\partial_2 v_\mathrm{f})v_\mathrm{f}^2.
\end{equation}
The vorticity,
\begin{equation}
  \omega_\mathrm{f} = \gamma_\odot^{-1}\partial_1v_\mathrm{f}^2 -
  \gamma_\odot^{-1}\partial_2(\gamma_\odot v_\mathrm{f}^1)
  = \gamma_\odot^{-1}\partial_1v_\mathrm{f}^2 - \partial_2v_\mathrm{f}^1 + \tau_\odot
 v_\mathrm{f}^1
\end{equation}	   
as it follows from its definition, $\omega := \smash{\lim_{\Delta
x^1\Delta x^2\to 0}} \smash{\frac{1}{\gamma_\odot\Delta x^1\Delta
x^2}}\smash{\oint} (\gamma_\odot v_\mathrm{f}^1\d{x^1} +
v_\mathrm{f}^2\d{x^2})$, and noting that
$\gamma_\odot'(x^2)/\gamma_\odot(x^2) = -\tau_\odot(x^2)$. 

A version of the Maxey--Riley equation that is suitable for geophysical
applications follows by replacing $\dot v_\mathrm{p}$ in the original
equation \eqref{eq:MR83} by \eqref{eq:a-f} with $v_\mathrm{f} =
v_\mathrm{p}$ and $\smash{\frac{\D{}}{\D{t}}}v_\mathrm{f}$ by
\eqref{eq:DvDt-f}, viz.,
\begin{equation}
  \dot v_\mathrm{p} +
  \left(\left(f_\mathrm{p}\!+\!\tfrac{1}{2}R\omega_\mathrm{f}\right)
  J\!+\!\frac{\Id}{\tau}\right)
  v_\mathrm{p} = \tfrac{3}{2}R\frac{\D{v_\mathrm{f}}}{\D{t}} +
  \left(\tfrac{3}{2}R\left(f_\mathrm{f}
  \!+\!\tfrac{1}{3}\omega_\mathrm{f}\right)
  J\!+\!\frac{\Id}{\tau}\right) v_\mathrm{f},
  \label{eq:MRgeo}
\end{equation}
where $f_{\boldsymbol{\cdot}} : = f + \tau_\odot v^1_{\boldsymbol{\cdot}}$.
A convenient simplification which treats $(x^1,x^2)$ \emph{as if}
it were Cartesian position as in our original setting is defined
by $\gamma_\odot = 1$, $\tau_\odot = 0$, and $f_{\boldsymbol{\cdot}}
= f = f_0 + \beta x^2$.  This is called a \emph{$\beta$-plane
approximation}, valid for $|x^2| \ll a_\odot$, with caveats
\cite{Ripa-JPO-97b}.

\begin{remark}
We will herein make use of the $\beta$-plane approximation for
simplicity of exposition, with $x^1$ (resp., $x^2$) Cartesian and
pointing eastward (resp., northward).  Results due to Coriolis
effects do not change when working on full spherical geometry.
\end{remark}

\begin{remark}
A version of the Maxey--Riley equation with Coriolis force appears in
\cite{Provenzale-99}. That version, however, as also includes the
centrifugal force, which is exactly balanced by the gravitational
force on a plane tangent to the Earth's surface.
\end{remark}

Application of nonautonomous GSPT analysis when $\tau = O(\varepsilon)$
as $\varepsilon\to 0$ leads to the following reduced equation on
the slow manifold:
\begin{equation}
  \dot x = v_\mathrm{p} = v_\mathrm{f} + \tau
  \big(\tfrac{3}{2}R-1\big)\left(\frac{\D{v_\mathrm{f}}}{\D{t}}
  + f Jv_\mathrm{f}\right)
  \label{eq:MRgeo-slow}
\end{equation}
$+\, O(\varepsilon^2)$.  Note the presence of the Coriolis term in
\eqref{eq:MRgeo-slow}, while the lift term makes an $O(\varepsilon^2)$
contribution to the slow manifold, as already noted above.  The
Coriolis term determines the behavior near geophysical vortices,
as we review next.  However, neither the lift term \emph{nor} the
Coriolis force contribute to set the convergence to, or divergence
away from, the neutral manifold \eqref{eq:N} as all the results
stated in Sec.\ \ref{sec:neutral} remain valid despite $A = -
\smash{\big(}\nabla v_\mathrm{f} +
(f+\smash{\tfrac{1}{3}}\omega_\mathrm{f}) J + \tau^{-1}\Id\smash{\big)}$
in \eqref{eq:doty} \citep[][App.\ B]{Beron-etal-19-PoF}. Remark
\ref{re:neutral}, which is expected to hold with the inclusion of
the Coriolis force, can be consequential for the interpretation of
the trajectories of (quasi) isopycnic and deep isobaric floats in
the ocean, which remain at the depths of prescribed density and
pressure surfaces, respectively.  The result on the local instability
of the slow manifold stated in Remark \ref{re:mono} also holds with
Coriolis force as $\Gamma = - \smash{\big(}\nabla v_\mathrm{f} +
(f+\smash{\tfrac{1}{2}}R\omega_\mathrm{f}) J + \tau^{-1}\Id\smash{\big)}
+ O(\varepsilon)$, which leads to $\Gamma + \Gamma^\top = -2(S +
\tau^{-1}\Id) + O(\varepsilon)$.  This shows that the local instability
of the slow manifold, not determined by the lift force, is not
influenced by the Coriolis force either.

\subsection{Inertial particle motion near geophysical vortices}

Motivated by astrophysical applications \cite{Tanga-etal-96},
Provenzale \cite{Provenzale-99} present results from numerical
simulations at low Rossby number (the relative-to-Coriolis
characteristic acceleration ratio \cite{Pedlosky-87}) suggesting
that the overall effect of the Coriolis force is to push heavy
particles toward the center of anticyclonic vortices (i.e., which
rotate against the local planet's spin sense).

Beron-Vera et al.\ \cite{Beron-etal-15} provided theoretical support,
in addition to numerical evidence, to a more general result about
behavior of inertial particles near quasigeostrophic (i.e.,
low-Rossby-number) eddies: anticyclonic/cyclonic eddies attract
(resp., repel) heavy/light (resp., light/heavy) particles.  This
result followed by first noting that a reduced Maxey--Riley equation
\eqref{eq:MRslow} consistent with quasigeostrophic flow, namely,
$\partial_t = O(\varepsilon)$, $v_\mathrm{f} = J\nabla\psi +
O(\varepsilon^2)$ and $f = f_0 + O(\varepsilon)$ (as $\varepsilon\to
0$, parameter that we are using to measure smallness throughout)
takes the form:
\begin{equation}
  \dot x = v_\mathrm{p} = J\nabla\psi +
  \tau\big(1-\tfrac{3}{2}R\big)f_0\nabla\psi \label{eq:MRgeo-slow-qg}
\end{equation}
$+\, O(\varepsilon^2)$. Quick inspection of \eqref{eq:MRgeo-slow-qg}
reveals that inertial effects should promote divergence away from,
or convergence into, \emph{Lagrangian eddies} whereas fluid particles
circulate around them without bypassing their boundaries.  By
``Lagrangian eddy'' we mean a vortex with a material boundary, i.e.,
composed of the same fluid particles, which is detected using an
objective (observer-independent) method \cite{Haller-Beron-13,
Haller-Beron-14, Haller-etal-16, Haller-etal-18}.  For a deeper
insight, let $U(t)\in D$ be a fluid region which is classified as
Lagrangian eddy at time $t$; let $\partial U(t)$ be its boundary.
The flux across $\partial U(t)$ \cite{Beron-etal-15}
\begin{equation}
   \mathcal F = \oint_{\partial U(t)} (\nabla\psi - Jv_\mathrm{p})\cdot
	\d{x} = \tau\big(1-\tfrac{3}{2}R\big)f_0\int_{U(t)}\nabla^2\psi\,
   \mathrm{d}^2x
\end{equation}
$+\, O(\varepsilon^2)$. Noting that $\nabla^2\psi$ is (the lowest-order
contribution in $\varepsilon$ to the) carrying flow vorticity, one
concludes that cyclonic ($f_0\nabla^2\psi > 0$) Lagrangian eddies
attract ($\mathcal F < 0$) light ($R > \smash{\frac{2}{3}}$) particles
and repel ($\mathcal F > 0$) heavy ($R < \smash{\frac{2}{3}}$)
particles, and vice versa for anticyclonic  ($f_0\nabla^2\psi < 0$)
eddies. This result confirms that for heavy particles obtained by
Provenzale \cite{Provenzale-99} based on numerical experimentation
and extends it for light particles.  For neutrally buoyant ($R =
\smash{\frac{2}{3}}$) particles $\mathcal F = 0$, just as if these
were fluid particles.

\begin{remark}
The above result is quite different than the nonrotating result,
in which case $\mathcal F = \tau(1-\smash{\frac{3}{2}}R)Q$ where
$Q := -\smash{\frac{\D{}}{\D{t}}}\nabla\cdot v_\mathrm{f} =
\smash{\frac{1}{2}}\smash{(\omega_\mathrm{f}^2 - S_\mathrm{s}^2 -
S_\mathrm{n}^2)}$.  Near the core of a Lagrangian vortex one
necessarily has $Q > 0$, which is the Okubo--Weiss criterion
\cite{Provenzale-99} (the condition $Q>0$, however, does not in
general guarantee the presence of a vortex due to the observer-dependence
of this diagnostic \cite{Haller-05, Beron-etal-13}).  The flux
criterion states that vortices attract light while repell heavy
particles, irrespective of their polarity.  \label{re:moha}
\end{remark}

A more rigorous statement of Beron-Vera et al.'s \cite{Beron-etal-15}
result can be made if the reference Lagrangian eddy is coherent in
the rotational sense of \citet{Haller-etal-16}.   To see how, let's
recall that a \emph{rotationally coherent eddy} (\emph{RCE}) is a
region $U(t)$, $t\in [t_0,t_0+T]$, enclosed by the outermost,
sufficiently convex isoline of the \emph{Lagrangian averaged vorticity
deviation} (\emph{LAVD}) enclosing a nondegenerate maximum.  For
QG flow, this objective quantity is given by
\begin{equation}
  \mathrm{LAVD}_{t_0}^t(x_0) := \int_{t_0}^t
  \big|\nabla^2\psi(F_{t_0}^s(x_0),s) -
  \overline{\nabla^2\psi}(s)\big| \d{s}.
\end{equation}
Here $F_{t_0}^t(x_0)$ is a trajectory of $J\nabla\psi$ starting
from $x_0$ at time $t_0$, and the overbar indicates average over
the fluid domain. Elements of $\partial U(t)$ complete the same total material
rotation relative to the mean material rotation of the whole mass
of fluid that contains it.  This property is observed \cite{Haller-etal-16}
to restrict the filamentation of $\partial U(t)$ to be mainly
tangential.  Let $\mathbf F_{t_0}^t(x_0)$ be the trajectory produced
by an arbitrary velocity field. By Liouville's theorem \cite{Arnold-89},
if
\begin{equation}
  \det\D{\mathbf F_{t_0}^t(x_0)} < 1
\end{equation}
over $[t_0,t_0+T]$, then  $\mathbf F_{t_0}^t(x_0)$ will be observably
\emph{attracting} over $[t_0,t_0+T]$.   Let $C(t_0)$ be the region
filled with closed isolines of $\mathrm{LAVD}_{t_0}^{t_0+T}(x_0)$
around 
\begin{equation}
  x_0^* = \argmax_{x_0\in C(t_0)}
  \mathrm{LAVD}_{t_0}^{t_0+T}(x_0).
\end{equation}
Consider an \emph{inertial} particle to be
$\varepsilon$-close to $x_0^*$ at time $t_0$, i.e., $x(t_0) = x_0^*
+ \varepsilon$.  By the smooth dependence of the solution of
\eqref{eq:MRgeo-slow-qg} on parameters, it follows that
\begin{equation}
  x(t;x(t_0),t_0) = x(t;x_0^*,t_0) + O(\varepsilon) =
  F_{t_0}^t(x_0^*) + O(\varepsilon).
  \label{eq:x0*}
\end{equation}
Now, take $\mathbf F_{t_0}^t(x_0) = x(t;x(t_0),t_0)$ and assume
that $D$ is sufficiently large for $ \overline{\nabla^2\psi}(t)
\approx 0$.  Using \eqref{eq:x0*} one finally obtains \cite{Haller-etal-16}
\begin{equation}
  \det\D{\mathbf F_{t_0}^{t_0+T}(x_0^*)} =
  \exp\tau\big(1 -
  \tfrac{3}{2}R\big)f_0\, 
  \mathrm{sLAVD}_{t_0}^{t_0+T}(x_0^*)
\end{equation}
$+\,O(\varepsilon^2)$ where
\begin{equation}
  \mathrm{sLAVD}_{t_0}^{t_0+T}(x_0^*) :=
  \!\!\!\sign_{t\in[t_0,t_0+T]}\!\!\!\nabla^2\psi(F_{t_0}^t(x_0^*),t),t)
  \,\mathrm{LAVD}_{t_0}^{t_0+T}(x_0^*), 
\end{equation}
from which one can state the following:
\begin{theorem}[Haller et al.\ \cite{Haller-etal-16}] 
The trajectory of the center of a cyclonic ($f_0\,\mathrm{sLAVD}>0$)
RCE is a finite-time attractor for light ($R>2/3$) particles, while
is a finite-time repellor for heavy ($R<2/3$) particles, and vice
versa for the trajectory of the center of an anticyclonic
($f_0\,\mathrm{sLAVD}<0$) RCE.  
\label{thm:h16}
\end{theorem}

\subsection{Observational support of the theory}

Beron-Vera et al. \cite{Beron-etal-15} present observational evidence
in support of the behavior predicted by Thm.\ \ref{thm:h16}.
Particularly revealing is the behavior described by two RAFOS floats
\cite{Wooding-etal-15} in the southeastern North Pacific.  RAFOS
floats are acoustically tracked buoys that are designed to drift
below the ocean surface along a preset nearly isobaric (depth)
level.

\begin{figure}[t!]
  \centering%
  \includegraphics[width=\textwidth]{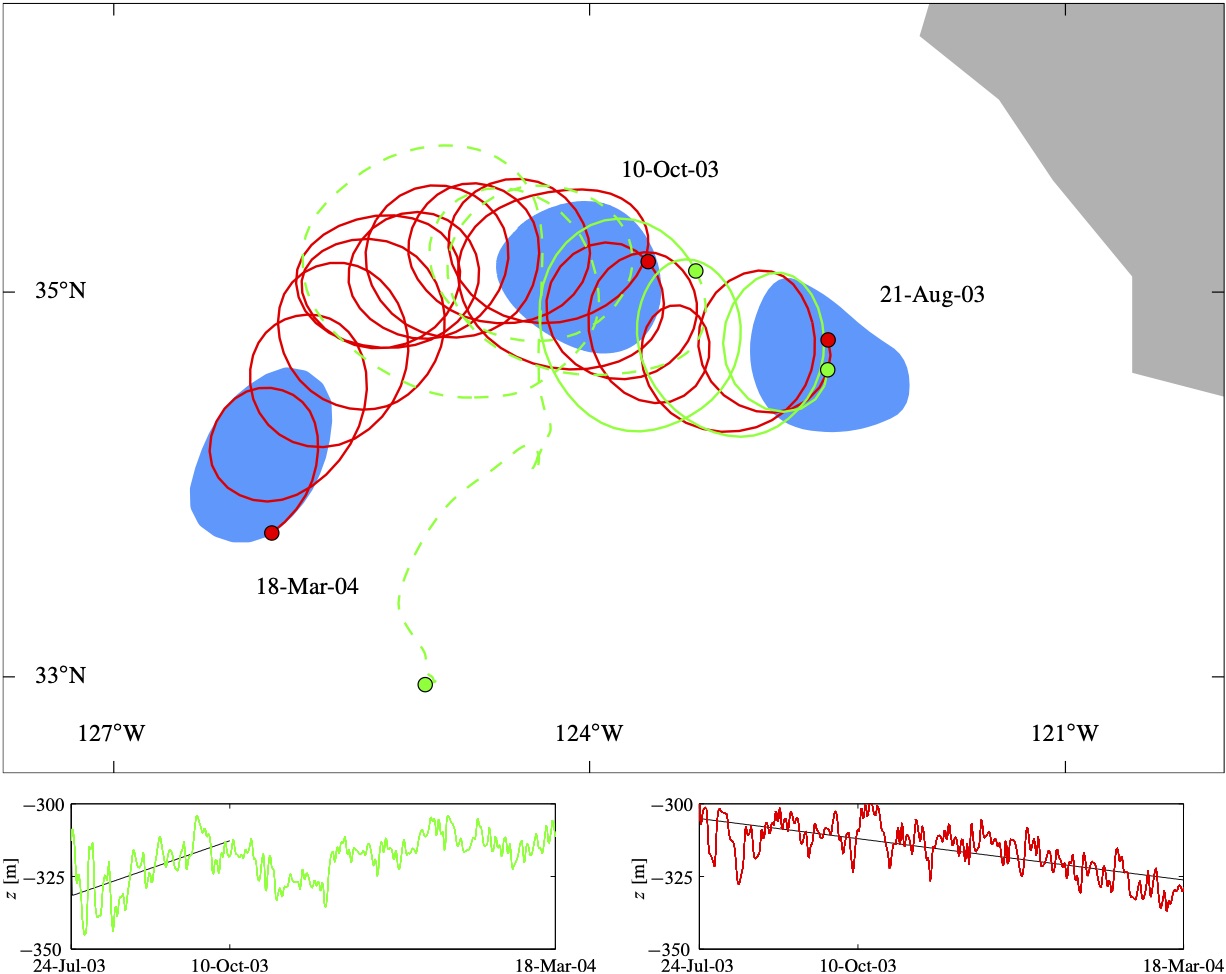}%
  \caption{ (top panel) Trajectories of two (acoustically tracked,
  submerged, quasi-isobaric) RAFOS floats (green and red) and
  snapshots of a California Undercurrent eddy or ``cuddy,'' detected
  from altimetry and classified as an RCE (light blue).  The dots
  indicate the positions of the floats on the dates that this
  anticyclonic mesoscale eddy is shown. (bottom-left panel) As a
  function of time, depth of the green float and that of an equivalent
  light particle under the action of gravity, buoyancy, and Stokes
  drag over the period in which the float remains inside the eddy
  (black).  (bottom-right panel) As in the right panel, but for the
  red float and a heavy particle.  Adapted from \citet{Beron-etal-15}.}
  \label{fig:rafos}%
\end{figure}

Initially close together, the two floats (indicated in red and green
in the top panel of Fig.\ \ref{fig:rafos}) were seen to take
significantly divergent trajectories on roughly the same depth level
(320 m).  This behavior at first glance might be attributed to
sensitive dependence of trajectories on initial positions in a
turbulent ocean.  But analysis of satellite altimetry measurements
of sea-surface height \cite{LeTraon-etal-98} reveals that the floats
on the date of closest proximity fall within a California Undercurrent
eddy or ``cuddy'' \cite{Garfield-etal-99} which is furthermore
classified as a coherent Lagrangian eddy \cite{Beron-etal-15}.
(Sea-surface height, $\eta$, represents a flow streamfunction
($\psi$) under the assumption of a quasigeostrophic balance between
the Coriolis force and the pressure gradient force, with the latter
resulting exclusively from differences in $\eta$
\citep[e.g.,][]{Beron-etal-08-GRL}.) However, while one float is
seen to loop anticyclonically accompanying this mesoscale eddy very
closely, the other float anticyclonically spirals away from the
eddy rather quickly (the portion of the trajectory when the float
is outside of the eddy is indicated in dashed).

The above seeming contradiction is resolved by noting that the green
float experiences a net ascending motion from 24 July 2003, the
beginning of the record, through about 10 October 2003, roughly
when the float escapes the coherent Lagrangian eddy, detected from
altimetry on 21 August 2003 (Fig.\ \ref{fig:rafos}, bottom-left
panel). By contrast, the red float indicated oscillates about a
constant depth over this period, but experiences a net descending
motion from 10 October 2003 until the end of the observational
record, 18 March 2004 (Fig.\ \ref{fig:rafos}, bottom-right panel).
Positive overall buoyancy can thus be inferred for the green float
from the beginning of the observational record until about 10 October
2003.  By contrast, negative overall buoyancy, preceded by a short
period of neutral overall buoyancy, can be inferred for the red
float over the entire observational record.

The sign of the overall buoyancy of each float can be used to
describe its behavior qualitatively using Thm.\ \ref{thm:h16}.  The
green float remains within the anticyclonic coherent Lagrangian
eddy from 21 August to around 10 October 2003, nearly when it leaves
the eddy and does not come back during the total observational
record (about 6 months).  This is qualitatively consistent with the
behavior of a \emph{light} particle.  Beyond 10 October 2003, the
buoyancy sign for this float is not relevant, given that it is
already outside the eddy.  In contrast, the red float remains inside
the Lagrangian eddy over the whole observational record.  This is
qualitatively consistent with the behavior of a \emph{heavy} particle.

The quantitative analysis in App.\ D of \citet{Beron-etal-15} shows
that $\delta - 1 \approx 0.5$ and $-0.1\times 10^{-10}$ for the
green (light) and red (heavy) floats, respectively, both characterized
by an inertial response time $\tau \approx 0.1$ d.  Taking $V =
0.1$ ms$^{-1}$ and $L = 50$ km as upper bounds on the tangential
speed and radius of cuddies \citep{Steinberg-etal-19}, respectively,
one gets $\smash{\frac{\tau}{L/V}} \approx 0.01$, which justifies
modeling the floats as inertial particles.

\begin{remark}  
Comparisons of theoretical predictions with additional observations
are presented in \cite{Beron-etal-15}.  These turned out to be
relatively less successful than the comparison just described.  The
main reason is the inability of the geophysically adapted Maxey--Riley
set to fully describe inertial ocean dynamics in the presence of
windage, which is the subject hereafter.
\end{remark}

\section{Maxey--Riley equation for surface ocean inertial dynamics}

The original Maxey--Riley equation and the geophysical adaptation
discussed above assume that the particles are \emph{immersed} in
the fluid.  This constrains the transferability of the latter to
ocean as it cannot fully describe the motion of \emph{floating
matter} such as marine debris of varied kinds
\cite{Trinanes-etal-16,Miron-etal-19-Chaos}.  This is mainly due
to its inability to simulate the effects of the \emph{combined
action of ocean currents and wind drag}.  These effects were accounted
for in a recent further adaptation of the Maxey--Riley equation to
oceanography by Beron-Vera, Olascoaga and Miron \cite{Beron-etal-19-PoF}.
The resulting equation, referred to as the \emph{BOM equation}, was
tested quite positively in the field
\cite{Olascoaga-etal-20,Miron-etal-20-GRL} as well as in the
laboratory \cite{Miron-etal-20-PoF}.

\subsection{The BOM equation}

\begin{figure}[t!]
  \centering%
  \includegraphics[width=\textwidth]{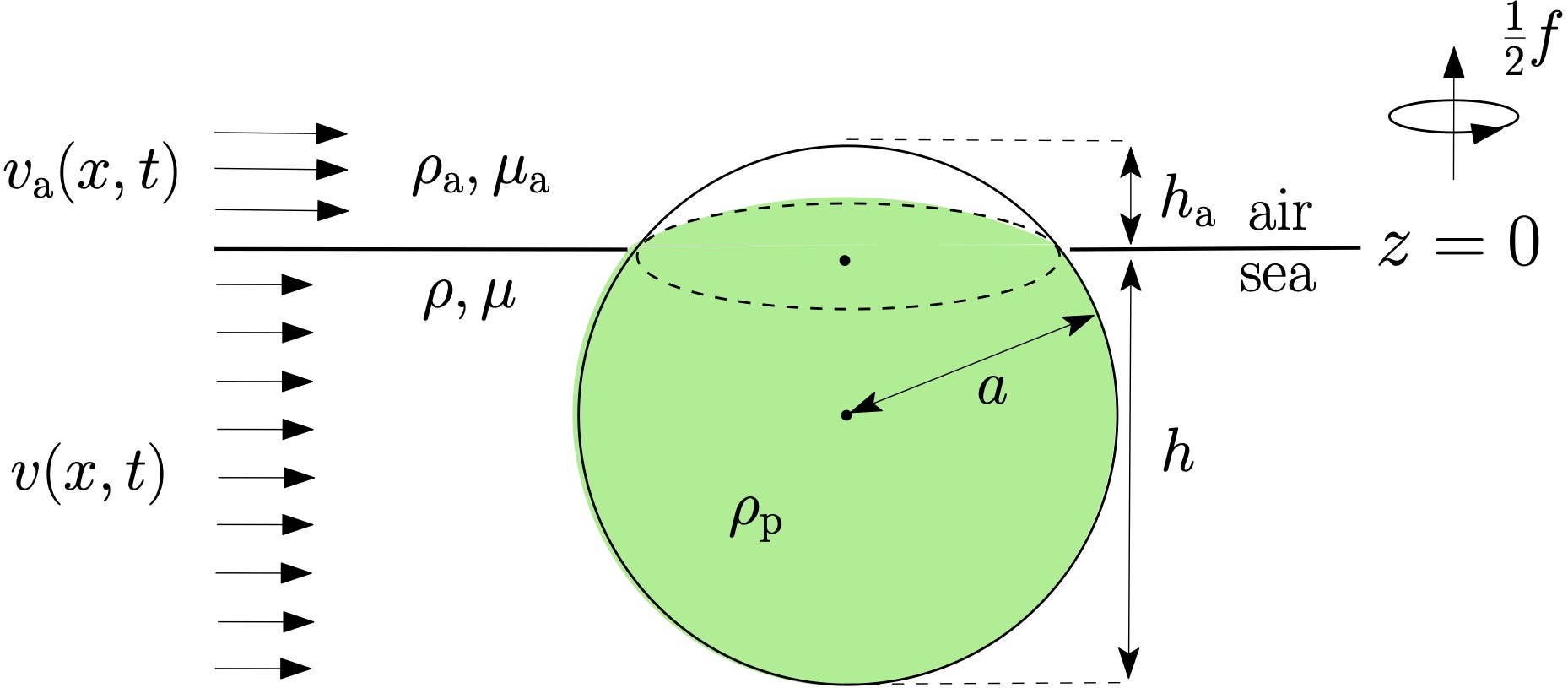}%
  \caption{Solid spherical particle that floats at an assumed flat
  interface between homogeneous seawater and air, and is subjected
  to flow, added mass, and drag forces resulting from the action
  of unsteady, horizontally sheared ocean currents and winds.  The
  various variables and parameters are defined in the text.
  Adapted from \citet{Beron-etal-19-PoF}.}
  \label{fig:bom}%
\end{figure}

Consider a stack of two homogeneous fluid layers separated by an
interface \emph{fixed} at $z = 0$  ($z$ is the vertical coordinate),
which rotates with angular speed $\frac{1}{2}f$, where $f (= f_0 +
\beta x^2$) is the Coriolis parameter  (Fig.\ \ref{fig:bom}).  The
fluid in the bottom layer represents the seawater and has density
$\rho$.  The top-layer fluid is much lighter, representing the air;
its density is $\rho_\mathrm{a} \ll \rho$.  Let $\mu$ and
$\mu_\mathrm{a}$ stand for dynamic viscosities of seawater and air,
respectively.  The seawater and air velocities vary in horizontal
position and time, and are denoted $v(x,t)$ and $v_\mathrm{a}(x,t)$,
respectively.  Consider finally a solid spherical particle (of
small radius $a$ and density $\rho_\mathrm{p}$) \emph{floating} at the
air--sea interface.

The exact fraction of submerged particle volume
\cite{Beron-etal-19-PoF,Olascoaga-etal-20}
\begin{equation}
  \sigma = \frac{1-\delta_\mathrm{a}}{\delta-\delta_\mathrm{a}},
  \label{eq:sga}
\end{equation}
where
\begin{equation}
  1\le \delta \le \frac{\rho}{\rho_\mathrm{a}} \gg 1,\quad
  \frac{\rho_\mathrm{a}}{\rho} 
  \le \delta_\mathrm{a} :=
  \frac{\rho_\mathrm{a}}{\rho_\mathrm{p}} \le 1,
\end{equation}
as static stability (Archimedes' principle) demands, so $0\le \sigma
\le 1$. The quantity $1-\sigma$ is sometimes referred to as
\emph{reserved volume}. Note that $\rho\gg\rho_\mathrm{a}$ implies
$\delta\gg\delta_\mathrm{a}$ and as a result $\sigma \approx
(1-\delta_\mathrm{a})/\delta$, which may be further approximated
by $\delta^{-1}$ \emph{if} $\delta_\mathrm{a} \ll 1$, which we
will assume herein.

\begin{remark}
It is important to realize that $\sigma \approx \delta^{-1}$ does
not follow from $\rho\ll \rho_\mathrm{a}$ as incorrectly stated in
\cite{Beron-etal-19-PoF}.  It is an assumption which holds
\emph{provided that $\delta$ is not too large}.  This follows from
noting that $\delta_\mathrm{a} \equiv (\rho_\mathrm{a}/\rho)\delta$.
Thus inferences made in \cite{Beron-etal-19-PoF} on behavior of the
BOM equation, presented below, as $\delta\to \infty$ are not formally
correct and should be ignored or interpreted with the these comments
in mind \cite{Olascoaga-etal-20}.
\end{remark}

\begin{remark}
The configuration in Fig.\ \ref{fig:bom} is susceptible to
(Kelvin--Helmholtz) instability \cite{LeBlond-Mysak-78}, which is
ignored assuming that the air--sea interface remains horizontal at
all times.  In other words, any wave-induced Stokes drift
\cite{Phillips-77} is accounted for implicitly, and admittedly only
partially, by absorbing its effects in the water velocity $v$ (e.g.,
as it would be directly measured or produced by some coupled
ocean--wave--atmosphere model).
\end{remark}

The emerged (resp., submerged) particle piece's height $h_\mathrm{a}$
(resp., $h = 2a - h_\mathrm{a}$) can be expressed in terms of
$\delta$ noting that
\begin{equation}
  h_\mathrm{a}/a = \Phi :=
  \smash{\frac{\mathrm{i}\sqrt{3}}{2}}(\varphi^{-1} - \varphi) -
  \frac{1}{2}(\varphi^{-1} + \varphi) + 1
\end{equation}
$\in [0,2)$ ,
where
\begin{equation}
  \varphi^3 := \mathrm{i}\sqrt{1 - (2\delta^{-1} - 1)^2} +
  2\delta^{-1} - 1.
\end{equation}
The emerged (resp., submerged) particle's projected (in the flow
direction) area $A_\mathrm{a}$ (resp., $A = \pi a^2 - A_\mathrm{a}$)
is also a function of $\delta$ since
\begin{equation}
  A_\mathrm{a}/\pi a^2 = \Psi := \pi^{-1}\cos^{-1}(1 - \Phi) - \pi^{-1}(1 - \Phi)
  \sqrt{1 - (1 - \Phi)^2}
\end{equation}
$ \in [0,1)$. 

Noting that fluid variables and parameters take different values
when pertaining to seawater or air, e.g.,
\begin{equation}
 v_\mathrm{f}(x,z,t) = 
  \begin{cases} 
	v_\mathrm{a}(x,t) & \text{if } z \in (0,h_\mathrm{a}], \\
   v(x,t)  & \text{if } z \in [-h,0),
  \end{cases}
\end{equation}
the BOM equation follows by vertically averaging each term of the
original Maxey--Riley equation, adapted to account for Earth’s rotation
effects, over the vertical extent $z\in [-h,h_\mathrm{a}]$ of the
particle.  The result is \cite{Beron-etal-19-PoF}
\begin{equation}
  \dot v_\mathrm{p} + \Big(\left(f + \tfrac{1}{3}R\omega\right)J +
  \frac{\Id}{\tau}\Big)
  v_\mathrm{p} = R\frac{\D{v}}{\D{t}} +
  R\left(f + \tfrac{1}{3}\omega\right)Jv +
  \frac{u}{\tau}, 
  \label{eq:BOM}
\end{equation}
where 
\begin{equation}
  u : = (1-\alpha) v + \alpha v_\mathrm{a}
  \label{eq:u}
\end{equation}
and $\smash{\frac{\D{}}{\D{t}}} v = \partial_t v + (\nabla v)v$
\citep[for a full spherical version of \eqref{eq:BOM}, cf.\ App.\
A of Ref.][]{Beron-etal-19-PoF}.

Primary BOM equation parameters $a$ and $\delta$ determine secondary
parameters $\alpha$, $R$, and $\tau$ as follows: 
\begin{equation}
  \alpha := \frac{\gamma\Psi}{1 + (\gamma - 1)\Psi}
  \label{eq:alpha}
\end{equation}
$\in [0,1)$,  which makes the convex combination \eqref{eq:u} a
weighted average of water and air velocities ($\gamma \approx 0.0167$
is the air-to-water viscosity ratio);
\begin{equation}
  R : = \frac{1 - \frac{1}{2}\Phi}{1 -\frac{1}{6}\Phi}
\end{equation}
$\in [0,1)$ and 
\begin{equation}
  \tau :=
  \frac{1-\frac{1}{6}\Phi}{\big(1 +
  (\gamma - 1)\Psi\big)\delta^4}\cdot
  \frac{a^2\rho}{3\mu}
  \label{eq:tau}
\end{equation}
$>0$, which measures the inertial response time of the medium to
a particle floating at the air--sea interface.

\begin{remark}
Note that parameters $R$ and $\tau$ of the BOM equations are different
than those involved in the original (and geophysically adapted) Maxey--Riley
equation(s).  The same symbols are used with no fear of confusion
so the structure of the BOM equation resembles as closely as possible
that of the original Maxey--Riley equation.
\end{remark}

\begin{remark}
In writing  \eqref{eq:alpha} and \eqref{eq:tau} we followed the
closure proposal made by \citet{Olascoaga-etal-20} to fully determine
parameters $R$ and $\tau$ in terms of the carrying fluid system
properties and inertial particle characteristics.  The original
formulation \cite{Beron-etal-19-PoF} of these parameters involved
projected length factors, $k$ and $k_\mathrm{a}$. These should
depend on how much the sphere is exposed to the air or immersed in
the water to account for the effect of the air--sea interface
(boundary) on the determination of the drag.  The closure proposal
in \cite{Olascoaga-etal-20}, $k = k_\mathrm{a} = \delta^{-3}$,
assures the air component of the carrying flow field to dominate
over the water component as the particle gets exposed to the air
while reducing differences with observations.  A stronger foundation
for this closure should be sought, possibly resorting on direct
numerical simulations of low-Reynolds-number flow around an spherical
cap of different heights.  To the best of our knowledge, a drag
coefficient formula for this specific setup is lacking.  An important
aspect that these simulations should account for is the effect of
the boundary on which the spherical cap rests on.  Recent efforts
in this direction are reported in \citep{Zeugin-etal-20}.
\end{remark}

\begin{remark}
The BOM equation was obtained assuming $\sigma \approx \delta^{-1}$.
A more correct way to formulate the BOM equation so it is valid for
all possible $\delta$ values is by using the exact form of $\sigma$,
as given in \eqref{eq:sga}.  This way the $\sigma\to 0$ (equivalently,
$\delta\to \infty$) limit is symmetric with respect to the $\sigma\to
1$ (equivalently, $\delta\to 1$) limit, as it can be expected.
Also, additional terms, involving air quantities should be included,
both in the full BOM equation and its reduced, slow-manifold
approximation, presented below, if $\delta$ is allowed to take
values in its full nominal range.  However, since $\delta < 10$ or
so, typically, these additional terms can be safely neglected.
\end{remark}

\begin{remark}
The weighted average of water and air velocities $u$ in \eqref{eq:u}
plays a very important role in short-term evolution
\citep{Olascoaga-etal-20, Miron-etal-20-GRL}, as we will review
below.  The velocity $u$ is of the type commonly discussed in the
search-and-rescue literature and referred to as ``leeway'' velocity
\cite{Breivik-etal-13}.  An important difference between \eqref{eq:u}
and the leeway modeling approach is that \eqref{eq:u} follows from
vertically averaging the drag force rather as an ad-hoc proposition
that involves an educated guess of leeway parameter $\alpha$
\citep[e.g.,][]{Trinanes-etal-16, Allshouse-etal-17}) or informed
by neglecting inertial effects and assuming an exact cancellation
of water and air (quadratic) drags \cite{Rohrs-etal-12, Nesterov-18},
which is at odds with the Maxey--Riley framework.
\label{rem:u}
\end{remark}

\begin{remark}
In deriving the BOM equation \eqref{eq:BOM} it is assumed that the
particle Reynolds number $\Rep$ \eqref{eq:rep} is small both above
and below the air--sea interface, which might thought difficult to
be satisfied given the difference in water and air speeds.  However,
results from field \citep{Olascoaga-etal-20, Miron-etal-20-GRL} and
laboratory \citep{Miron-etal-20-PoF} experiments, reviewed below,
justify the assumption of a Stokes regime in the water and the air.
This provides support to the earlier suggestion \citep{Beron-etal-19-PoF}
that an appropriate way to define $\Rep$ is using $V_\mathrm{slip}
= |v_\mathrm{p}-u|$, which is $O(\varepsilon)$ for sufficiently
small particles, fulfilling the small $\Rep$ requirement.
\end{remark}

\subsection{Slow-manifold approximation}

Assume that both $v(x,t)$ and $v_\mathrm{a}(x,t)$ are smooth in
each of their arguments.  In the limit when $\tau = O(\varepsilon)$
as $\varepsilon \to 0$, the BOM equation \eqref{eq:BOM} involves
both slow ($x$) and fast ($v_\mathrm{p}$) variables, which makes it
a singular perturbation problem, just as the Maxey--Riley equation
\eqref{eq:MR83} and its geophysical adaptation \eqref{eq:MRgeo}
under a similar assumption.  In these circumstances one can apply
nonautonomous GSPT to obtain the following reduced equation
\cite{Beron-etal-19-PoF}:
\begin{equation}
  \dot{x} = v_\mathrm{p} = u +  u_\tau
 \label{eq:BOMslow}
\end{equation}
$+\,O(\varepsilon^2)$, where
\begin{equation}
  u_\tau :=
  \tau \left(R\frac{\D{v}}{\D{t}} + R \Big(f +
  \frac{1}{3}\omega\Big) Jv - \frac{\D{u}}{\D{t}} -
  \Big(f + \frac{1}{3}R\omega\Big) Ju\right)
  \label{eq:utau}
\end{equation}
with $\smash{\frac{\D{}}{\D{t}}}u = \partial_t u + (\nabla u)u$,
i.e., the total derivative of $u$ \emph{along a trajectory of} $u$.

The reduced equation \eqref{eq:BOMslow} controls the evolution of the
full equation \eqref{eq:BOM} on the slow manifold, defined by
\begin{equation}
  M_\tau := \big\{(x,v_\mathrm{p},t) : v_\mathrm{p} = u(x,t) +
  u_\tau(x,t)\big\}.
  \label{eq:Mtaubom}
\end{equation}
Being $C^rO(\varepsilon)$-close to the critical manifold, given
by
\begin{equation}
  M_0 := \big\{(x,v_\mathrm{p},t) : v_\mathrm{p} = u(x,t)\big\},
  \label{eq:M0bom}
\end{equation}
for any $r<\infty$, and unique up to an error much smaller than
$O(\varepsilon)$, $M_\tau$ is a locally invariant, normally hyperbolic
manifold that attracts all solutions of \eqref{eq:BOM} exponentially
fast.  The only caveat \cite{Haller-Sapsis-08} is that rapid changes
in the carrying flow velocity, \emph{represented by} $u$, can turn
the exponentially dominated convergence of solutions on $M_\tau$
not necessarily monotonic over finite time.

\begin{remark}
The carrying flow ($u$) that defines the critical manifold $M_0$
depends on the buoyancy of the particle and thus has \emph{inertial
effects built in}. Inertial effects are felt by the particle even
during the initial stages of the evolution, which are controlled
by $\dot x = u$ provided that $v_p$ initially at $t=t_0$ is
$O(\varepsilon)$-close to $u$, as it follows from the smooth
dependence of the solutions of \eqref{eq:BOM} on parameters.  This
is important in comparisons with field and laboratory observations,
which we discuss below after illustrating long-time asymptotic
aspects of the BOM equation.
\end{remark} 

\begin{remark}
For neutrally buoyant particles ($\delta = 1$) the BOM equation
reduces \emph{exactly} to \eqref{eq:doty} except that $A = -
\smash{\big(}\nabla v + (f+\smash{\tfrac{1}{3}}\omega) J +
\tau^{-1}\Id\smash{\big)}$.  However, all the results stated in
Sec.\ \ref{sec:neutral} relating to the stability of the neutral
manifold $N$ \eqref{eq:N} hold \cite{Beron-etal-19-PoF}.  An important
observation is that, unlike in the original Maxey--Riley equation
\eqref{eq:MR83} and its geophysical adaptation \eqref{eq:MRgeo},
$N$ does not coincide with the critical manifold $M_0$ \eqref{eq:M0bom}.
\end{remark}

\subsection{Local instability of the slow manifold}

Applying the local instability analysis of \cite{Haller-Sapsis-10},
discussed in Remark \ref{re:nile}, on the slow manifold of the BOM
equation \eqref{eq:Mtaubom}, one finds that perturbations off it will
grow where
\begin{equation}
  \tau > \frac{2}{\sqrt{(\partial_1u^1-\partial_2u^2)^2 +
  (\partial_2u^1+\partial_1u^2)^2} - \nabla\cdot u}.
\end{equation}
This new result follows upon noting that $\Gamma = - \smash{\big(}\nabla
u + (f+\smash{\tfrac{1}{3}}R\omega) J + \tau^{-1}\Id\smash{\big)}
+ O(\varepsilon)$, which leads to $\Gamma + \Gamma^\top = - \nabla
u - (\nabla u)^\top - 2\tau^{-1}\Id + O(\varepsilon)$.  Note that
neither the lift term nor the Coriolis force contribute to set the
instability of the slow manifold, a property of the Maxey--Riley
equation \eqref{eq:MR83} and its geophysical adaptation \eqref{eq:MRgeo}.
The practical consequence of the result just presented awaits to
be investigated.

\subsection{Behavior near quasigeostrophic eddies}

Theorem \ref{thm:h16}, though successful in describing the behavior
of submerged floats, falls short at explaining an observed
\cite{Brach-etal-18} tendency of floating plastic debris to collect
inside anticyclonic mesoscale eddies while avoiding cyclonic ones.
The BOM equation turns out to be capable of describing this
observation, as articulated next.

Oceanic mesoscale eddies (with diameters ranging from 50 to 250 km)
are characterized by a low Rossby number \cite{Pedlosky-87}, so it
is reasonable to explore the local stability of floating inertial
particles near the center of quasigeostrophic RCE as done to arrive
at Thm.\ \ref{thm:h16}.  The starting point is the reduced BOM
equation \eqref{eq:BOMslow}, approximated by
\begin{equation}
  \dot x = v_\mathrm{p} = gf_0^{-1}J\nabla\eta +
  \tau g(1-\alpha-R)\nabla\eta
  \label{eq:BOMslow-qg}
\end{equation}
$+\,O(\varepsilon^2)$.  This approximations holds under the following
assumptions. First, $v = gf_0^{-1}J\nabla\eta + O(\varepsilon^2)$,
where $\eta(x,t)$ is sea surface height and $g$ stands for gravity,
$\partial_t = O(\varepsilon)$, and $f = f_0 + O(\varepsilon)$.
Second $\alpha = O(\varepsilon)$, at least, consistent with it being
very small (a few percent) over a large range of buoyancy ($\delta$)
values. Third, $v_\mathrm{a} = O(\varepsilon^2)$, at least, i.e.,
the wind field over the period of interest is sufficiently weak
(calm).

Applying on \eqref{eq:BOMslow-qg} the same local stability
analysis that led to Thm.\ \ref{thm:h16}, one finds
\cite{Beron-etal-19-PoF}
\begin{equation}
  \det\D{\mathbf F_{t_0}^{t_0+T}(x_0^*)} =
  \exp\tau\big(1 - R - \alpha\big)f_0\, 
  \mathrm{sLAVD}_{t_0}^{t_0+T}(x_0^*)
\end{equation}
$+\,O(\varepsilon^2)$ where
\begin{equation}
  \mathrm{sLAVD}_{t_0}^{t_0+T}(x_0^*) :=
  \!\!\!\sign_{t\in[t_0,t_0+T]}\!\!\!gf_0^{-1}\nabla^2\psi(F_{t_0}^t(x_0^*),t),t)
  \,\mathrm{LAVD}_{t_0}^{t_0+T}(x_0^*). 
\end{equation}
Since $1 - R \ge \alpha \ge 0$, on can state the following:
\begin{theorem}[Beron-Vera, Miron and Olascoaga \cite{Beron-etal-19-PoF}] 
The trajectory of the center of an anticyclonic ($f_0\,\mathrm{sLAVD}<0$)
RCE is a finite-time attractor for floating inertial particles,
while that of a cyclonic ($f_0\,\mathrm{sLAVD}>0$) RCE is a finite-time
repellor for floating inertial particles. 
\label{thm:bom19}
\end{theorem}

\subsection{Great garbage patches}

The ocean's subtropical gyres are well-documented
\cite{Cozar-etal-14,Lebreton-etal-18} to show a tendency to accumulate
plastic debris forming large patches, particularly that of the North
Pacific, known as the ``Great Pacific Garbage Patch.'' This tendency
of floating matter to concentrate in the subtropical gyres has been
noted \cite{Beron-etal-16} in the distribution of \emph{undrogued}
surface drifting buoys from the NOAA Global Drifter Program
\cite{Lumpkin-Pazos-07}.  A standard drifter from this program,
which collects data since 1979, follows the Surface Velocity Program
or SVP \cite{Niiler-etal-87} design with a 15-m-long holey-sock
drogue attached to it to minimize wind slippage and wave-induced
drift, thereby maximizing its water tracking characteristics.
However, the drogue is often lost after some time from deployment
\cite{Lumpkin-etal-12} while the satellite tracker included in the
spherical float keeps transmitting positions.

\begin{figure}[t!]
  \centering%
  \includegraphics[width=\textwidth]{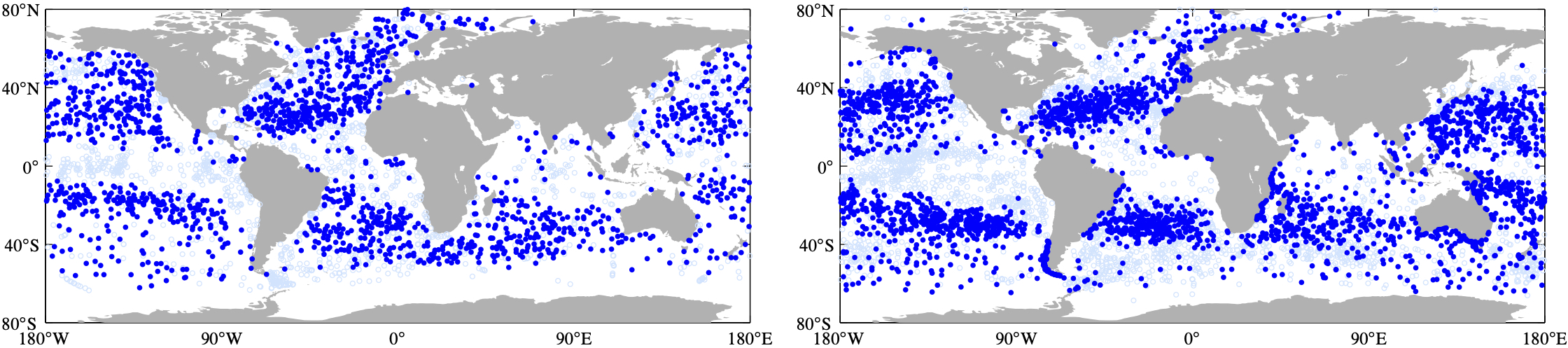}%
  \caption{Initial (light blue) and final (blue) positions of drogued
  (left) and undrogued (right) SVP drifters from the NOAA Global
  Drifter Program over 1979--present.  ``Final position'' refers
  to the last recorded position after at least 1 yr past the time
  at the ``initial position,'' which is the deployment position for
  drogued drifters or the location where a drifter loses the drogue.}
  \label{fig:svp}%
\end{figure}

The left panel of Fig.\ \ref{fig:svp} shows positions at deployment
time (light blue) and positions after a period of at least 1 yr
(blue) of all SVP drifters that remained drogued over the entire
period.  The right panel shows positions where the drifters have
lost their drogues (light blue) and positions taken by these drifters
after at least 1 yr from those instances (blue).  The initial
positions are similarly homogeneously distributed.  But there is a
marked difference in the final positions: while the drogued drifters
take a more homogeneous distribution, the undrogued drifters reveal
a tendency to accumulate in the subtropical gyres.

\begin{figure}[t!]
  \centering%
  \includegraphics[width=\textwidth]{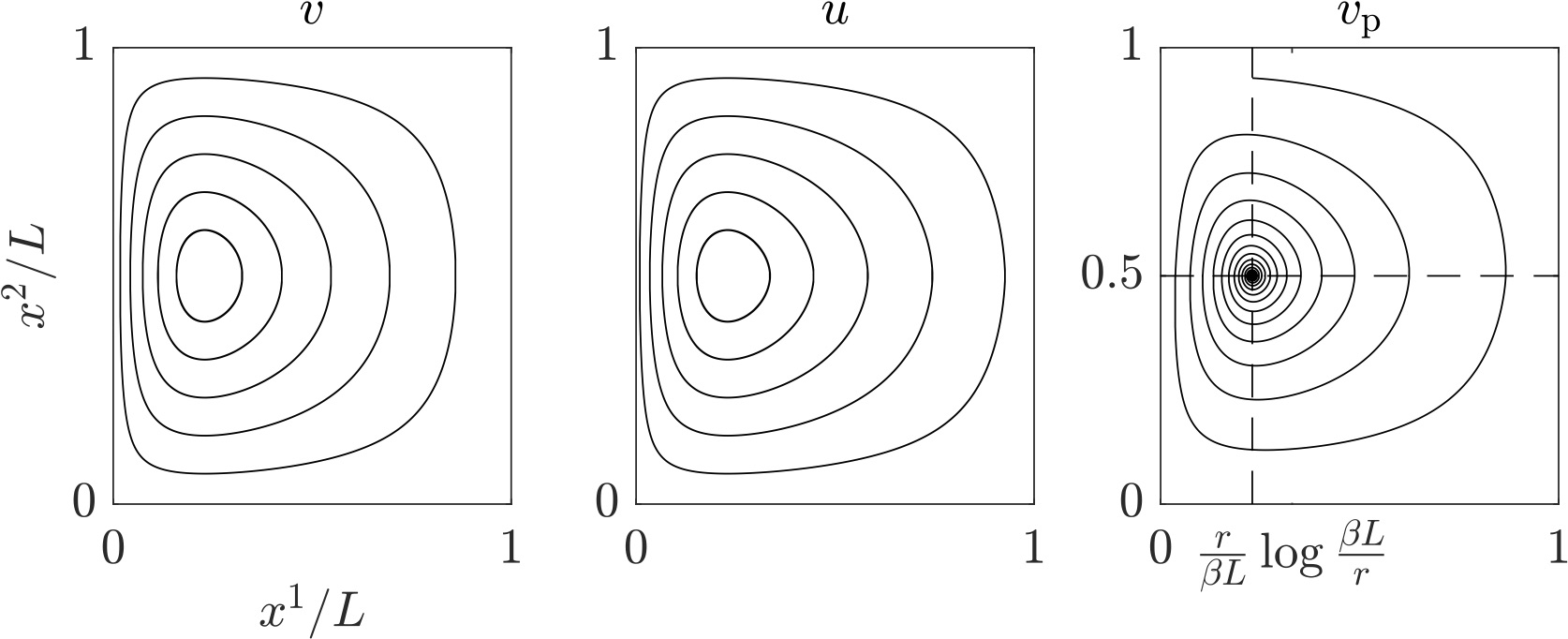}%
  \caption{Streamlines of the Stommel wind-driven circulation model
  velocity (left), the $\delta$-weighted-average of this velocity
  and the wind field that drives the Stommel gyre (middle), and the
  inertial particle velocity on the slow manifold of the BOM equation
  resulting from using these water and air velocities (right).
  Adapted from \citet{Beron-etal-19-PoF}.}
  \label{fig:stommel}%
\end{figure}

The BOM equation is able to predict great garbage patches in the
long run consistent with observed behavior, thereby allowing to
interpret this behavior as produced by inertial effects.  To see
this one can consider Stommel's \cite{Stommel-48} conceptual model
of wind-driven circulation as in \cite{Beron-etal-19-PoF}.  The
steady flow in such a barotropic (constant density) model is
quasigeostrophic, i.e., $v = J\nabla\psi + O(\varepsilon^2)$, and
has an anticyclonic basin-wide gyre in the northern hemisphere, so
$\omega = \nabla^2\psi \le 0$, driven by steady westerlies and trade
winds, $v_\mathrm{a} = W(x^2)e_1$ with $W'(x^2) \ge 0$.  The inertial
particle velocity on the slow manifold \eqref{eq:BOMslow} takes the
form
\begin{equation}
  v_\mathrm{p} = (1-\alpha)J\nabla\psi + \alpha We_1 +
 \tau f_0 \big((1-R-\alpha) \nabla\psi - \alpha W e_2\big)
 \label{eq:stommel}
\end{equation}
with an $O(\varepsilon^2)$ error.  The divergence of this
velocity is given by
\begin{equation}
  \nabla\cdot  v_\mathrm{p} = \tau f_0 \big(
  (1-R-\alpha)\nabla^2\psi - \alpha W'(x^2)\big).
  \label{eq:div}
\end{equation}
Recalling that $1- R -\alpha \ge 0$, it follows that $\nabla\cdot
v_\mathrm{p} \le 0$, which promotes clustering of inertial particles
in the interior of the gyre in a manner akin to undrogued drifters
and plastic debris.  Moreover, in \cite{Beron-etal-19-PoF} it is
shown that $\dot x = v_\mathrm{p}$, with $v_\mathrm{p}$ as in
\eqref{eq:stommel} with $\psi$ as given in \cite{Haidvogel-Bryan-92}
and $W$ deduced from the wind stress using a bulk formula, has a
stable spiral equilibrium at $(x^1,x^2) = (\smash{\frac{r}{\beta}}
\log\smash{\frac{\beta L}{r}}, \smash{\frac{1}{2}}L)$ where $L$ is
the side of an assumed square midlatitude domain and $r$ is the
bottom friction coefficient.  The right panel of Fig.\ \ref{fig:stommel}
shows streamlines of $v_\mathrm{p}$ assuming $a = 17.5$ cm and
$\delta = 2$ (which give $\alpha \approx 0.01$, $R \approx 0.6$,
and $\tau \approx 0.01$ d, roughly characterizing undrogued drifters).
Additional parameter choices, $H = 200$ m (thermocline depth), $L
= 10$ Mm, $r = 10^{-5}$ s$^{-1}$, $F = 2\times10^{-3}$ m$^2$s$^{-2}$
(wind stress amplitude per unit density), and $C_\mathrm{D} = 1.2
\times 10^{-3}$ (drag coefficient).  Note that a ``leeway'' model,
i.e., one of the form $\dot x = u$, produces closed streamlines
(Fig.\ \ref{fig:stommel}, middle panel) just as the Stommel model
streamlines (Fig.\ \ref{fig:stommel}, left panel).

\begin{figure}[t!]
  \centering%
  \includegraphics[width=\textwidth]{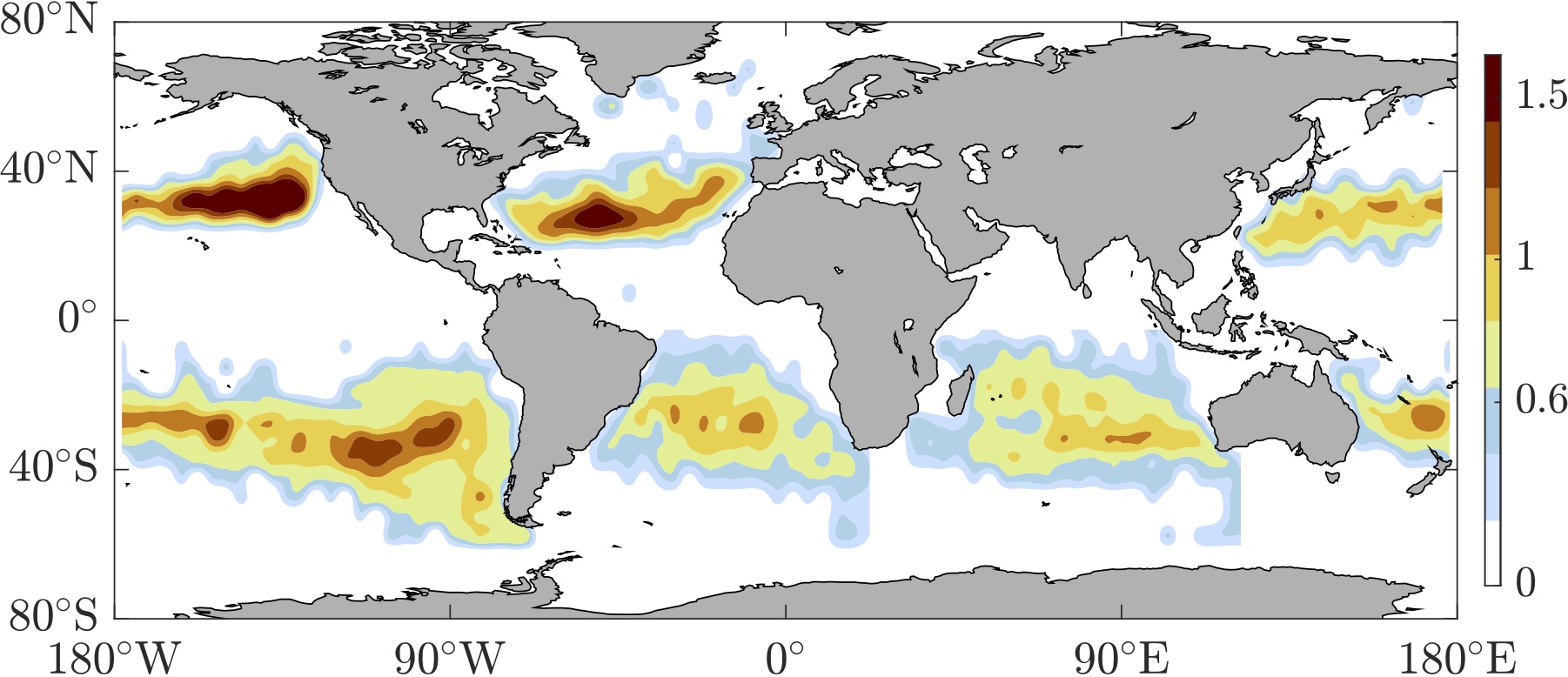}%
  \caption{Expressed as number per degree squared, density of
  inertial particles after 1.5 yr normalized by density in the
  initially uniform distribution of particles over the global surface
  ocean.  The water velocity is given by surface ocean velocity
  output from the Hybrid-Coordinate Ocean Model.  The air velocity
  corresponds to the wind velocity from the reanalysis used to
  construct the wind stress that forces the model. Adapted from
  \citet{Beron-etal-16}.}
  \label{fig:hycom}%
\end{figure}

Earlier studies have argued that the formation of great garbage
patches in the subtropical gyres is due to wind-induced downwelling
in such regions.  This is not represented in Stommel's model for
being a higher order (in the Rossby number) effect. Beron-Vera,
Olascoaga and Lumpkin \cite{Beron-etal-16} showed that clustering
of undrogued drifters in the subtropical gyres (Fig.\ \ref{fig:hycom}),
which is visible already after about 1.5 yr, is too fast to be
explained by wind-induced downwelling.  This was done by comparing
the long-term evolution of trajectories of $\dot x  = v$,  with $v$
produced by a \emph{general ocean circulation model}, with that of
floating inertial particles evolving under an earlier form of the
BOM equation.  Such an earlier form the BOM equation was obtained
by modeling the submerged (resp., emerged) particle portion as a
sphere of the fractional volume that is submerged (resp., emerged)
while it evolves under the geophysically adapted Maxey--Riley
equation.  Despite this earlier form of the BOM equation was
successful in explaining great garbage patch formation, it could
not explain the observed tendency of anticyclonic eddies to trap
plastic debris, which motivated the derivation of its successor.

\subsection{Field experiments verification}\label{sec:field}

Olascoaga et al.\ \cite{Olascoaga-etal-20} present results from a
field experiment that provides support to the BOM equation. The
field experiment consisted in deploying simultaneously specially
designed drifters of varied sizes, buoyancies, and shapes off the
southeastern Florida Peninsula in the Florida Current, and subsequently
tracking them via satellite. Four types of special drifters were
involved in the experiment, mimicking debris found in the ocean.
The main bodies of these special drifters represented a sphere of
radius 12 cm, approximately, a cube of about 25 cm side, and a
cuboid of approximate dimensions 30 cm $\times$ 30 cm $\times$ 10
cm.  These special drifters were submerged below the sea level by
roughly 10, 6.5, and 5 cm, respectively.  The fourth special drifter
consisted in an artificial boxwood hedge of about 250 cm $\times$
50 cm and thickness of nearly 2 cm.  It floated on the surface with
the majority of its body slightly above the surface.

To account for the effects produced by the lack of sphericity of
the special drifters, a simple heuristic fix, expected to be valid
for sufficiently small objects, was used consisting in multiplying
$\tau$ in \eqref{eq:tau} by $K$, a \emph{shape factor} satisfying
\cite{Ganser-93}
\begin{equation}
  K = \frac{3a_\mathrm{v}}{a_\mathrm{n} +
  2a_\mathrm{s}}.
  \label{eq:K}
\end{equation}
Here $a_\mathrm{n}$, $a_\mathrm{s}$, and $a_\mathrm{v}$ are the
radii of the sphere with equivalent projected area, surface area,
and equivalent volume, respectively, whose average provide an
appropriate choice for $a$.

\begin{table}[t!]
  \centering
  \begin{tabular}{lcccccccc}%
    \hline\hline\\[-.8em]%
	 && \multicolumn{7}{c}{Parameter}\\
	 \cline{3-9}\\[-.8em]
	 &&  \multicolumn{3}{c}{Primary} & & 
	 \multicolumn{3}{c}{Secondary}\\
	 \cline{3-5}\cline{7-9}\\[-.8em]
	 Drifter type && $a$ [cm] & $K$ & $\delta$ & & $\alpha$ & $R$ & $\tau$ [d]\\\\[-.8em]
	 \hline\\[-.8em]
	 Sphere &&  12 & 1.00 & 2.7 & & 0.027 & 0.51 & 0.002\\
	 Cube  && 16 & 0.96 & 4.0 & & 0.042 & 0.42 & 0.001\\
	 Cuboid && 13 & 0.95 & 2.5 & & 0.024 & 0.53 & 0.003\\
	 Hedge && 26 & 0.53 & 1.3 & & 0.005 & 0.79 & 0.031\\\\[-.8em]
    \hline%
  \end{tabular}%
  \caption{Parameters that characterize the special drifters as
  inertial particles.}%
  \label{tab:parameters}
\end{table}

The various parameters that characterize the special drifters as
inertial ``particles'' are shown in Table \ref{tab:parameters}.  An
a-priori dimensional analysis justifies treating them as such and
thus using the BOM equation to investigate their motion. Let $V$
and $L$ be typical carrying fluid system velocity and length scales,
respectively.  Taking $V = 1$ m\,s$^{-1}$, typical at the axis of
the Florida Current, and $L = 50$ km, a rough measure of the width
of the current, one obtains that $\smash{\frac{\tau}{L/V}} \ll 1$
for the special drifters as required since $L/V$ is of the order
of 1 d and $\tau$ is much shorter than that for the drifters.


\begin{figure}[ht!]
  \centering%
  \includegraphics[width=.75\textwidth]{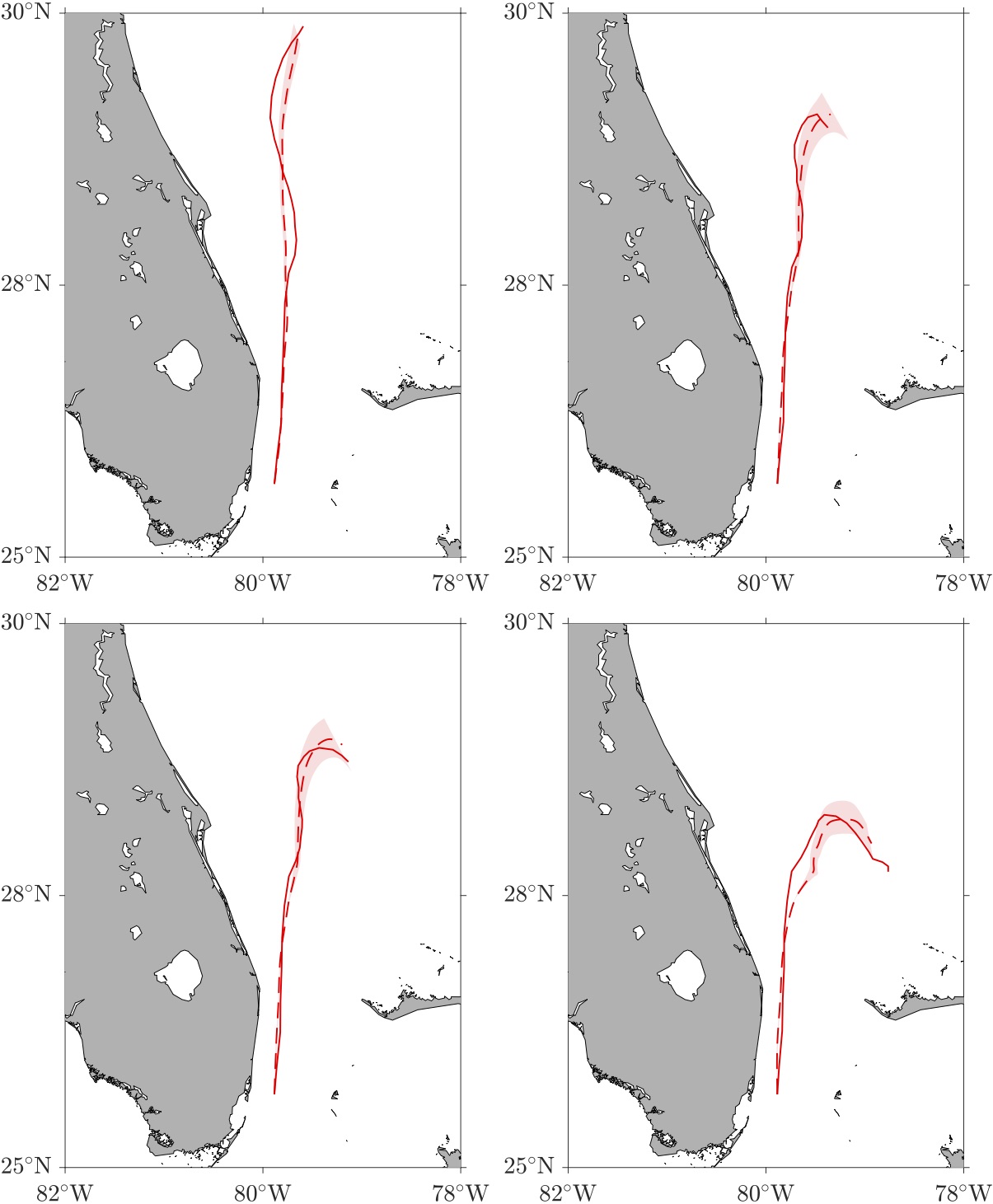}%
  \caption{One-week-long trajectories of specially designed undrogued
  drifters with buoyancy increasing rightward (solid) along with
  trajectories produced by BOM equation (dashed; shades reflect
  uncertainty around drifter buoyancy determination). Adapted
  from \citet{Olascoaga-etal-20}.}
  \label{fig:special}%
\end{figure}

Figure \ref{fig:special} shows week-long trajectories taken by the
special drifters (solid curves) along with trajectories produced
by the BOM equation (dashed curves with shades of red around
reflecting an assumed 10\pct\, uncertainty in the determination of
the buoyancy of the drifters).  The driving ocean currents are
provided by an altimetry/wind/drifter data synthesis
\cite{Olascoaga-etal-20} and the winds are from reanalysis
\cite{Dee-etal-11}.  The special drifter trajectories were subjected
to a strong wind event 2 to 3 days after deployment, which affected
them differently mainly according to their buoyancy as described
by the BOM equation.  We note that $\tau$ turned out to be sufficiently
small for the trajectories of the BOM equation \eqref{eq:BOM},
initialized from the special drifter deployment locations and
velocities estimated by differentiating the special drifter
trajectories, to be well approximated by those of $\dot x = u$ over
the initial stages of the evolution.

\begin{remark}
Indeed, let $x(t;x_0,t_0)$ denote the trajectory of a particle
starting from $x_0$ at time $t_0$.  By smooth dependence of the
solutions of the BOM equation \eqref{eq:BOM} on parameters, if the
particle is initialized with velocity $u(x_0,t_0)$, then $v_\mathrm{p}$
will remain $O(\varepsilon)$-close to $u(x(t;x_0,t_0),t)$ over
$[t_0,t]$ \emph{finite}.  In other words, over finite $[t_0,t]$ the
trajectory of the particle will be mainly controlled by the integrated
effect of the ocean current and wind drag, explaining why BOM
equation trajectories in Fig.\ \ref{fig:special} were well approximated
by those of $\dot x = u$.
\end{remark}

\begin{remark}
Further support for the validity of the BOM equation is provided
by Miron et al.\ \cite{Miron-etal-20-GRL}, who considered a much
larger set of longer special drifter trajectories, starting from
several locations in the tropical North Atlantic. Since various
special drifters of the same type were included in each deployment,
a cluster analysis was possible to be carried out showing grouping of
trajectories depending on drifter design.  This added further support
to the importance of inertial effects on floating matter drift.
Special drifter trajectories and BOM equation trajectories in many
cases showed very good agreement.  As in \cite{Olascoaga-etal-20}
the latter were seen to be well approximated by those of $\dot x =
u$ despite their longer extent (one month or longer vs.\ one week).
Disagreements were mainly attributed to limitations of the carrying
flow system representation as assessed by the low skill of the ocean
current representation in describing the motion of drogued drifters,
also included in the experiments.
\end{remark}

\subsection{Laboratory verification}

\begin{figure}[t!]
  \centering%
  \includegraphics[width=.75\textwidth]{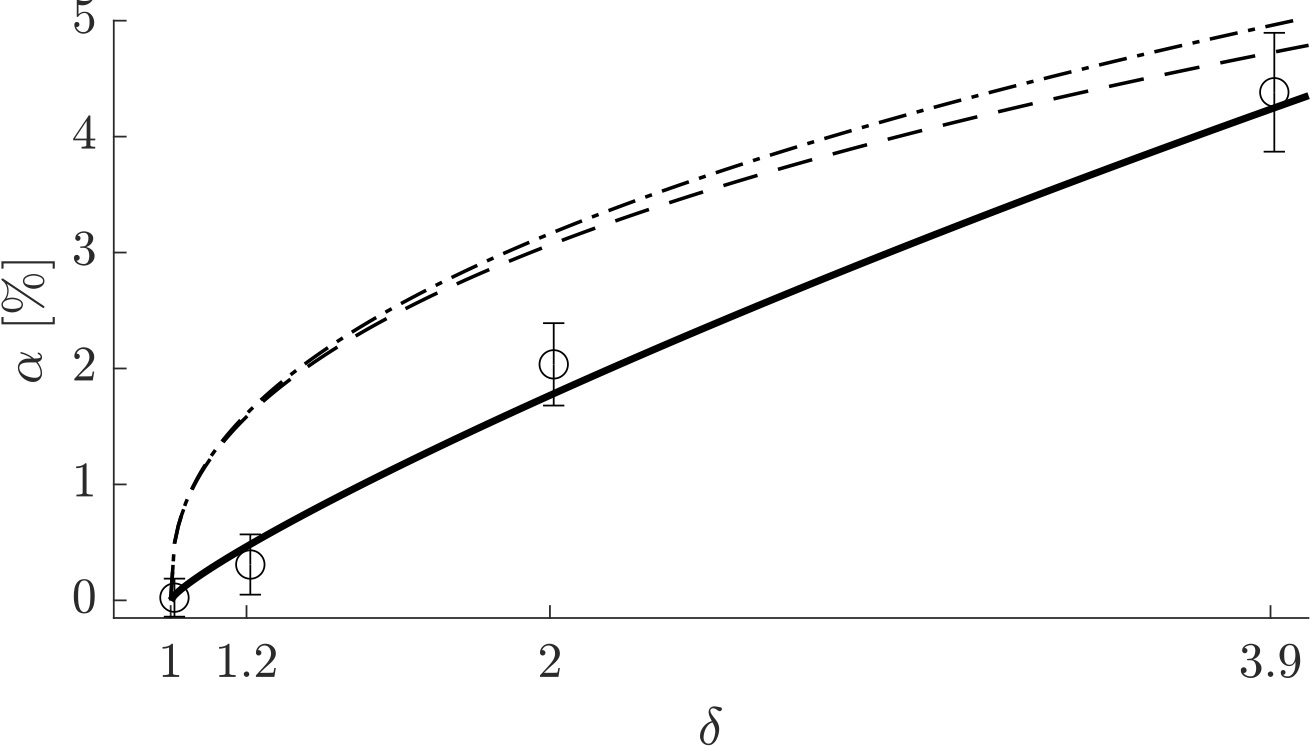}%
  \caption{As a function of buoyancy, estimated (circles) and
  theoretical (solid curve) ``leeway'' factor. The accompanying
  error bar represent one standard deviation uncertainties.  The
  dashed and dot-dashed curves are buoyancy-dependent ``leeway''
  models derived in \citep{Nesterov-18} and \citep{Rohrs-etal-12},
  respectively. Adapted from \citet{Miron-etal-20-PoF}.}
  \label{fig:tank}%
\end{figure}

Miron et al.\ \cite{Miron-etal-20-PoF} report results from a series
of experiments in an air--water stream flume facility that provide
controlled observational support for the buoyancy dependence of the
BOM equation's carrying flow velocity. This was found to play a
very important role in the field experiments just described, despite
the rough estimates of the buoyancy of the drifters in the tropical
North Atlantic experiments and the admittedly poor representations
of the carrying ocean currents and winds were available, both in
the Florida Current and tropical North Atlantic experiments.

The laboratory experiments were designed to specifically validate
the dependence of the ``leeway'' factor $\alpha$ on $\delta$ in
\eqref{eq:alpha}.  This was done by noting that when $v$ and
$v_\mathrm{a}$ are constant, and hence $u$ as well, in the nonrotating
case,
\begin{equation}
  v_\mathrm{p}(t) = v_\mathrm{p}(0)\mathrm{e}^{-t/\tau} +
  (1-\alpha)v + \alpha v_\mathrm{a}
  \label{eq:vp}
\end{equation}
\emph{exactly} solves the BOM equation \eqref{eq:BOM}, which allows
us to estimate $\alpha$ as a function of $\delta$ given $v$ and
$v_\mathrm{a}$ and measurements of $v_\mathrm{p}$ in the along-flume
direction.

The laboratory experiments were carried out \cite{Medina-20} in the
Air-Sea Interaction Salt-water Tank (ASIST) of the Alfred G.\
Glassell, Jr.\ SUrge STructure Atmosphere INteraction (SUSTAIN)
facility of the University of Miami's Rosenstiel School of Marine
\& Atmospheric Science (https://sustain.rsmas.miami.edu/).  ASIST
offers the possibility to control the water stream with a pump and
the air stream using a fan.  

Four thick rubber, deformation resistant balloons of equal radius
$a = 0.11$ m were employed in the experiments.  These were filled
with different water levels so that $\delta \approx$ 3.9, 2, 1.2,
and 1, which represent a fairly range of $\delta$ values given that
the corresponding submerged-depth-to-diameter ratios are $h/2a
\approx 0.33$, 0.5, 0.75, and 1.

The circles in Fig.\ \ref{fig:tank} are mean (over several experiment
realizations) $\alpha$ values estimated from \eqref{eq:vp} taking
$v$ as the vertical average of the water stream profile (estimated
using particle image velocimetry \cite{Novelli-etal-17}) over a
balloon diameter (the balloon velocities were estimated from video
tracking).  Error bars represent one standard deviation uncertainties.
Note the very good agreement with the theoretical $\alpha(\delta)$
curve \eqref{eq:alpha}, shown as solid curve.  Indeed, the agreement
between measurements and the BOM equation prediction is much better
than that between measurements and two buoyancy-dependent ``leeway''
parameter models discussed in the search-and-rescue literature
\cite{Rohrs-etal-12, Nesterov-18}, included for reference as dashed
and dot-dashed curves, respectively.

\begin{remark}
The laboratory experiment results suggest that neglecting the
Basset--Boussinesq history or memory term is indeed well justified
despite being of the same order as the drag term.  How this result
is altered by flow unsteadiness is not known and should be
investigated.\label{re:tel}
\end{remark}

\section{Maxey--Riley equation for elastically coupled floating particles}

One additional extension of the Maxey--Riley equation is presented
by Beron-Vera and Miron \cite{Beron-Miron-20}.  This is motivated
by an interest to understand the mechanism that leads \emph{Sargassum}
(a type of large brown seaweed) to inundate coastal waters and land
on beaches, particularly those of the Caribbean Sea.  This phenomenon
has been on the rise since 2011 \cite{Wang-etal-19,Johns-etal-20}
and is challenging scientists, coastal resource managers, and
administrators at local and regional levels \cite{Langin-18}.

\subsection{The \emph{Sargassum} drift model}

A raft of pelagic \emph{Sargassum} is composed of flexible stems
which are kept afloat by means of bladders filled with gas while
it drifts under the action of ocean currents and winds.   Beron-Vera
and Miron \cite{Beron-Miron-20} proposed a mathematical model for
this physical depiction of a drifting \emph{Sargassum} raft as an
elastic network of buoyant, finite-size particles that evolve
according to the BOM equation.

To construct the mathematical model, Beron-Vera and Miron
\cite{Beron-Miron-20} consider a network of $N > 1$ spherical
particles (beads) connected by (massless, nonbendable) springs.
The (small) particles are assumed to have $\delta \ge 1$ finite.
The elastic force (per unit mass) exerted on particle $i$, with
two-dimensional Cartesian position $x_i = (x^1_i,x^2_i)$, by neighboring
particles at positions $\{x_j : j\in \neigh(i)\}$, is assumed to
obey Hooke's law (e.g., \cite{Goldstein-81}):
\begin{equation}
F_i = 
- \sum_{j\in \neigh(i)} k_{ij}\big(|x_{ij}| -
\ell_{ij}\big)\frac{x_{ij}}{|x_{ij}|},
\label{eq:F}
\end{equation}
$i = 1,\dotsc,N$, where 
\begin{equation}
  x_{ij} := x_i - x_j;
  \label{eq:xij}
\end{equation}
$k_{ij} \ge 0$ is the stiffness (per unit mass) of the spring
connecting particle $i$ with neighboring particle $j$; and $\ell_{ij}
\ge 0$ is the length of the latter at rest. 

The \emph{Sargassum} drift model is obtained by adding the elastic
force \eqref{eq:F} to the right-hand-side of the BOM equation.  The
result is a set of $N$ 2nd-order ordinary differential equations,
\emph{coupled} by the elastic term, viz.,
\begin{equation}
  \dot v_i + \left(\left. f\right\vert_i +
  \tfrac{1}{3}R\left.\omega \right\vert_i\right)v_i^\perp +
  \frac{v_i}{\tau} = R\frac{\D{\left. v\right\vert_i}}{\D{t}} +
  R\left(\left. f\right\vert_i +
  \tfrac{1}{3}\left.\omega\right\vert_i\right)\left
  .v\right\vert_i^\perp + \frac{\left.
  u\right\vert_i}{\tau} + F_i,
  \label{eq:SM}
\end{equation}
$i = 1,\dotsc,N$, where $v_i$ is the velocity of particle $i$ and
$\left. \right\vert_i$ means pertaining to particle $i$.

Because the elastic force \eqref{eq:F} does not depend on velocity,
the nonautonomous GSPT analysis of the BOM equation with $\tau =
O(\varepsilon)$ as $\varepsilon\to 0$ \cite{Beron-etal-19-PoF}
applies to \eqref{eq:SM} with the only difference that the equations
on the slow manifold are coupled by the elastic force \eqref{eq:F},
namely,
\begin{equation}
 \dot x_i = v_i = \left. u\right\vert_i +  \left.
 u_\tau\right\vert_i + \tau F_i + O(\tau^2),  
 \label{eq:SMslow}
\end{equation}
$i = 1,\dotsc, N$. The slow manifold of \eqref{eq:SM} is the $(2N
+ 1)$-dimensional subset $\{(x_i,v_i,t) : v_i = u(x_i,t) + u_\tau(x_i,t)
+ \tau F_i(x_i; x_j:j\in \neigh(i)) + O(\tau^2),\, i = 1,
\dotsc, N\}$ of the $(4N + 1)$-dimensional phase space $(x_i,v_i,t)$,
$i = 1, \dotsc, N$.

\subsection{Behavior near quasigesotrophic eddies}

Equation \eqref{eq:SMslow}, which attracts all solutions of
\eqref{eq:SM}, can be approximated by
\begin{equation}
 \dot x_i = v_i = gf_0^{-1}\nabla^\perp\left.\eta\right\vert_i +
 \tau \big(g(1-\alpha-R)\nabla\left.\eta\right\vert_i + F_i\big)
 \label{eq:SMslow-qg}
 \end{equation}
$+\,O(\varepsilon^2)$, $i = 1,\dotsc, N$, under the following
assumptions. First, the near surface ocean flow is in quasigeostrophic
balance, i.e., $v = gf_0^{-1}\nabla^\perp\eta + O(\varepsilon^2)$,
$\partial_t = O(\varepsilon)$, and $f = f_0 + O(\varepsilon)$.
Second,  the elastic interaction does not alter the nature of the
critical and slow manifolds, which is guaranteed by making $F_i =
O(\varepsilon)$.  Third, $\alpha = O(\varepsilon)$, at least,
consistent with it being very small (a few percent) over a large
range of buoyancy ($\delta$) values. Fourth, $v_\mathrm{a} =
O(\varepsilon^2)$, at least, i.e., the wind field over the period
of interest is sufficiently weak (calm).

Applying a local stability analysis similar to the one applied on
\eqref{eq:MRgeo-slow-qg} and \eqref{eq:BOMslow-qg}, one obtains the following:
\begin{theorem}[Beron-Vera and Miron \cite{Beron-Miron-20}]
The trajectory of the center of an RCE, $F_{t_0}^{t}(x_0^*)$, is
locally forward attracting overall over $t\in [t_0,t_0+T]$:
\begin{enumerate}
  \item for all $k_{ij}$ when $\sign_{t\in
  [t_0,t_0+T]}\nabla^2\eta(F_{t_0}^t(x_0^*),t) < 0$; and
  \item provided that
   \begin{equation}
     |T|\sum_{i=1}^N\sum_{j\in \neigh(i)} k_{ij} >
	  gN(1-\alpha-R)
	  \left\vert\int_{t_0}^{t_0+T}
     |\nabla^2\eta(F_{t_0}^t(x_0^*),t)|\d{t}\right\vert 
     \label{eq:k}
   \end{equation}
   when $\sign_{t\in
  [t_0,t_0+T]}\nabla^2\eta(F_{t_0}^t(x_0^*),t) > 0$.
\end{enumerate}
\label{thm:bm20}
\end{theorem}
Since $\omega = gf_0^{-1}\nabla^2\eta + O(\varepsilon^2)$, the above
result says that the center of a cyclonic rotationally coherent
quasigeostrophic eddy represents a finite-time attractor for elastic
networks of inertial particles in the presence of calm winds if
they are sufficiently stiff, while that of an anticyclonic eddy
irrespective of how stiff.

Table 

\begin{table}[t!]
  \centering
  \begin{tabular}{lllc} 
  \hline\hline\\[-0.75em]
  Type of inertial particle(s)				  & Cyclonic vortices & Anticyclonic vortices & Thm.  \\\\[-0.75em]%
  \hline\\[-.9em]%
  Submerged, light                          & Attract           & Repel                 & 1     \\%
  Submerged, heavy                          & Repel             & Attract               & 1     \\%
  Floating, free                            & Repel             & Attract               & 2     \\%
  Floating, elastic network, weakly stiff   & Repel             & Attract               & 3     \\%
  Floating, elastic network, strongly stiff & Attract           & Attract               & 3     \\\\[-.9em]%
  \hline
  \end{tabular}
  \caption{Summary of the results from Thms.\
  \ref{thm:h16}--\ref{thm:bm20}.}%
  \label{tab:t13}
\end{table}

\subsection{Reality check}

\begin{figure}[t!]
  \centering%
  \includegraphics[width=\textwidth]{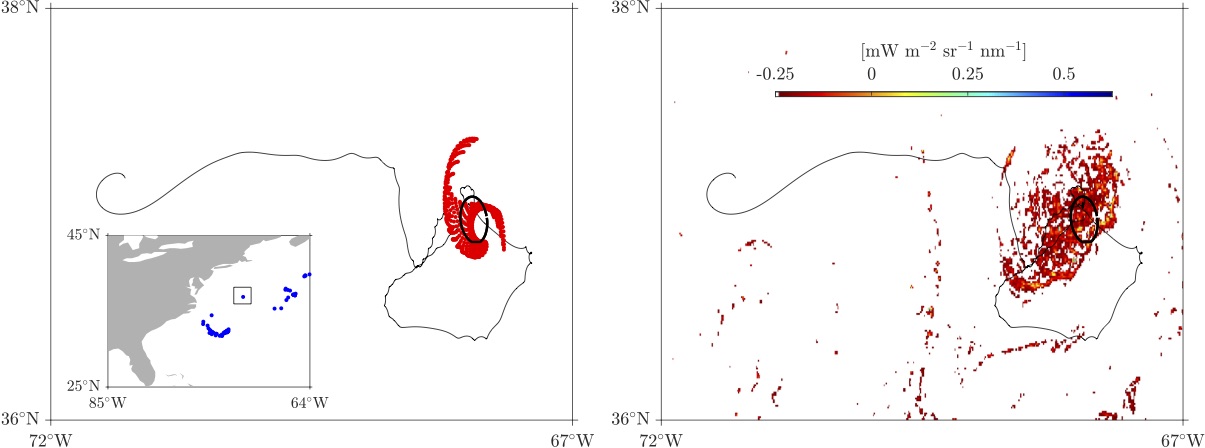}%
  \caption{(left panel) Elastically interacting inertial particles
  (red) concentrating inside a cold-core (i.e., cyclonic) Gulf
  Stream ring (with boundary depicted in thick black and trajectory
  indicated by the thin black curve) compared to noninteracting
  particles that are repelled away from the ring (blue).  The ring
  was inferred from altimetry and classified as an RCE.  The
  particles, whose positions are shown on 7 October 2006, were
  initiated 6 months earlier from exactly the same locations near
  the boundary of the ring. (right panel) The boundary of the ring
  on the left with observed \emph{Sargassum} concentrating within
  (\emph{Sargassum} corresponds to Maximum Chlorophyll Index (MCI)
  values exceeding $-0.25$\,mW\,m$^{-2}$sr$^{-1}$nm$^{-1}$).}
  \label{fig:sar}%
\end{figure}

The predictions of Thm.\ \ref{thm:bm20} are consistent with
observations, as is exemplified in Fig.\ \ref{fig:sar}. Blue dots
in the left panel are noninteracting inertial particles, while red
dots are inertial particles connected elastically. The particles,
whose positions are shown on 7 October 2006, were initiated 6 months
earlier from exactly the same locations near the boundary of a
cold-core (i.e., cyclonic) Gulf Stream ring. Detected from altimetry
and classified as an RCE, the boundary of the ring is depicted in
thick black.  Its trajectory (since 10 April 2006) is indicated by
the thin black curve.  In the simulation, the network's springs are
taken of equal length at rest, $\ell_{ij} = 0.5$ m.  The beads,
totalling $n = 625$, have a common radius $a = 0.1$ m.  The buoyancies
of the beads are all taken the same and equal to $\delta = 1.25$,
following Olascoaga et al.\ \cite{Olascoaga-etal-20}.  The resulting
inertial parameters $\alpha = 5.9\times 10^{-3}$, $R = 0.6$, and
$\tau = 4.1\times 1^{-2}$.

Note the effect of the ring on the elastically interacting particles.
This is in stark contrast with that on the noninteracting inertial
particles, which are repelled way away from the ring.  Indeed, on
7 October 2006 the noninteracting particles lie more than 1000 km
away from the ring.  The concentration of the elastically interacting
particles inside the cyclonic RCE, predicted by Thm.\ \ref{thm:bm20}.2,
is consistent with the observation, shown in the right panel of
Fig.\ \ref{fig:sar}, of \emph{Sargassum} as inferred from MODIS
(Moderate Resolution Imaging Spectroradiometer) satellite imagery.
Note that the concentration of \emph{Sargassum} is high in the Gulf
Stream ring in question.

\begin{remark}
An important observation is that the elastically interacting particles
have been evolved under the full system \eqref{eq:SM} rather than
\eqref{eq:SMslow-qg}, which approximates the reduced system
\eqref{eq:SMslow} assuming quasigeostrophic ocean currents and calm
wind conditions.  While the ocean currents were inferred from
altimetry, which is consistent with the quasigeostrophic assumption,
the wind was provided by reanalysis data with no restriction of any
kind on its intensity.  This suggests that Thm.\ \ref{thm:bm20} is
valid on a wider range of conditions than formally required.
\end{remark}

\begin{remark}
The oceanographic relevance of the results above is that eddies,
observed to propagate westward \cite{Morrow-etal-04a} consistent
with theoretical expectation \cite{Cushman-etal-90}, can provide
an effective mechanism for the connectivity of \emph{Sargassum}
between the Intra-Americas Sea and remote regions in the
tropical/equatorial Atlantic.
\end{remark}

\section{Concluding remarks}

Despite the significant progress already made in recent years to
port the Maxey--Riley framework to oceanography, a number of aspects
still need to be accounted for to expand its applicability. For
instance, at present wave-induced drift effects are represented
implicitly in the BOM equation, at the carrying flow system level.
Explicit representation of these effects, whose importance awaits
to be carefully assessed, should account for the tendency of waves
to push objects downward when they are close to the air--sea
interface, which might be parametrized by making the object's
buoyancy a function of the angle of wave attack.  This would require
controlled experimentation in a wind-wave tank facility.  Shape
effects are at the moment represented heuristically.  Direct
computational fluid dynamics experimentation would be needed to
derive appropriate formulas for the drag depending on the object's
shape. Sinking and rising of plastic debris as well as \emph{Sargassum}
rafts are reported. This would require one to include a buoyancy
force.  Clearly, in this case a reliable representation of
three-dimensional ocean currents would be critical.  Physiological
changes of \emph{Sargassum} are necessary to be accounted to enable
a more accurate description of the evolution of rafts.  This should
minimally control the growth and decay of size of the elastic
networks as they drift across regions of the ocean with varying
thermal and geochemical conditions.  The practical utility of the
BOM equation is not restricted to marine debris and \emph{Sargassum}
raft motion prediction. Among the many additional problems that the
BOM equation should be useful for are search-and-rescue operations
at sea and the drift of sea-ice in a warming climate.   In every
case the nonlinear dynamics techniques and results overviewed here
once appropriately adapted are expected to facilitate the understanding
of observed behavior as well as predicting behavior yet to be
observed.

\begin{acknowledgements}
  The constructive criticism of two anonymous reviewers led to
  improvements to this paper.  I want to acknowledge the influence
  exerted by Gustavo Go\~ni on my career for triggering my interest
  in nonlinear dynamics while I was an undergraduate student of
  oceanography, and by Paco Villaverde, Roberto Delellis, the late
  Pedro Ripa, and Mike Brown for conveying subsequent sustain to
  it.  Doron Nof's presentation at the 2016 Ocean Science Meeting
  \cite{Nof-16} provided inspiration for the work that led to the
  derivation of the BOM equation in collaboration with Maria Olascoaga
  and Philippe Miron, with whom I am in debt for the benefit of
  many discussions on inertial ocean dynamics.  I thank George
  Haller and Chris Jones for clarifying comments on geometric
  singular perturbation theory.  Remark \ref{re:moha} is due to
  Mohammad Farazmand.  Remark \ref{re:tel} was brought to my attention
  by Tam\'as T\'el.  This work builds on lectures I imparted at the
  CISM--ECCOMAS International Summer School on ``Coherent Structures
  in Unsteady Flows: Mathematical and Computational Methods,''
  Udine, Italy, 3--7 June 2019, organized by George Haller.  Support
  for the work overviewed here was provided by CONACyT--SENER
  (Mexico) grant 201441, the Gulf of Mexico Research Initiative,
  and the University of Miami's Cooperative Institute for Marine
  and Atmospheric Science.
\end{acknowledgements}

\section*{Conflict of interest}

The author declares that he has no conflict of interest.

\bibliographystyle{spbasic} 

\end{document}